\documentclass[aps, prb, reprint, amsmath, amssymb, groupedaddress, 
nolongbibliography]{revtex4-2}
\usepackage[colorlinks=true, allcolors=blue]{hyperref}
\usepackage{graphicx}
\usepackage{overpic}
\usepackage{mathrsfs}
\newcommand{\lt}{\left}
\newcommand{\rt}{\right}
\newcommand{\pa}{\partial}
\newcommand{\bx}{\mathbf{x}}

\newcommand{\bk}{\mathbf{k}}
\newcommand{\br}{\mathbf{r}}
\newcommand{\bh}{\mathbf{h}}
\newcommand{\bH}{\mathbf{H}}
\newcommand{\cS}{\mathcal{S}}
\newcommand{\Gr}{\mathcal{G}}

\newcommand{\lambdar}{\tilde{\lambda}}
\newcommand{\mur}{\tilde{\mu}}
\newcommand{\Tr}{\tilde{T}}
\newcommand{\sigmar}{\tilde{\sigma}}
\newcommand{\Cr}{\tilde{C}}
\newcommand{\Sr}{\tilde{S}}
\newcommand{\alphar}{\tilde{\alpha}}
\newcommand{\Yr}{\tilde{Y}}
\newcommand{\rhor}{\tilde{\rho}}
\newcommand{\kappar}{\tilde{\kappa}}
\newcommand{\Br}{\tilde{B}}

\begin{document}
\title{Perturbative renormalization and thermodynamics of quantum crystalline membranes}
\author{Achille Mauri}
\email[]{a.mauri@science.ru.nl}

\author{Mikhail I. Katsnelson}
\affiliation{Radboud University, Institute for Molecules and Materials, Heyendaalseweg 
135, 6525 AJ Nijmegen, The Netherlands}

\date{\today}

\begin{abstract}
We analyze the statistical mechanics of a free-standing quantum crystalline membrane 
within the framework of a systematic perturbative renormalization group (RG).
A power-counting analysis shows that the leading singularities of correlation functions 
can be analyzed within an effective renormalizable model in which the kinetic energy of 
in-plane phonons and subleading geometrical nonlinearities in the expansion of the strain 
tensor are neglected.
For membranes at zero temperature, governed by zero-point motion, the RG equations of the 
effective model provide a systematic derivation of logarithmic corrections to the bending 
rigidity and to the elastic Young modulus derived in earlier investigations.
In the limit of a weakly applied external tension, the stress-strain relation at $T = 0$ 
is anomalous: the linear Hooke's law is replaced with a singular law exhibiting 
logarithmic corrections.
For small, but finite temperatures, we use techniques of finite-size scaling to derive 
general relations between the zero-temperature RG flow and scaling laws of thermodynamic 
quantities such as the thermal expansion coefficient $\alpha$, the entropy $S$, and the 
specific heat $C$.
A combination of the scaling relations with an analysis of thermal fluctuations shows 
that, for small temperatures, the thermal expansion coefficient $\alpha$ is negative and 
logarithmically dependent on $T$, as predicted in an earlier work.
Although the requirement $\lim_{T \to 0} \alpha = 0$, expected from the third law of 
thermodynamics is formally satisfied, $\alpha$ is predicted to exhibit such a slow 
variation to remain practically constant down to unaccessibly small temperatures.
\end{abstract}

\maketitle

\section{Introduction}

The statistical mechanics of fluctuating elastic membranes has been investigated 
extensively over the last decades, in connection with a broad variety of physical systems, 
from biological layers to graphene and other atomically-thin two-dimensional 
materials.
As it has long been realized, models of flexible surfaces subject to vanishing or small 
external tension exhibit a rich and striking physical behavior, controlled by an interplay 
between fluctuations and mechanical nonlinearities~\cite{nelson_statistical, 
nelson_jpf_1987, david_epl_1988, aronovitz_prl_1988, aronovitz_jpf_1989, guitter_jpf_1989, 
katsnelson_graphene, kownacki_pre_2009, kosmrlj_prb_2016, los_prl_2016, 
gornyi_2dmater_2017, los_npj_2017, le-doussal_aop_2018, saykin_prr_2020, coquand_pre_2020, 
coquand_pre_2021, shankar_pre_2021, metayer_pre_2022, pikelner_arxiv_2021}.
A crucial prediction, in particular, is that for homogeneous free-standing membranes 
without an imposed stress, anharmonicities inherent in the geometrical definition of the 
elastic strain tensor are responsible for the stabilization of a macroscopically flat 
phase at finite temperatures~\cite{nelson_statistical, nelson_jpf_1987, david_epl_1988, 
aronovitz_prl_1988}, and for a dramatic power-law renormalization of the effective 
scale-dependent bending rigidity and elastic constants.
The presence of a quenched disorder, besides thermal fluctuations, has been predicted to 
induce an even richer physical behavior~\cite{nelson_statistical, gornyi_2dmater_2017, 
le-doussal_aop_2018, saykin_prr_2020, coquand_pre_2021}.
Renormalizations of elastic and thermodynamic properties by ripples have also been 
analyzed in experiments on graphene membranes~\cite{lopez-polin_np_2015, 
blees_nature_2015, nicholl_prl_2017, lopez-polin_nms_2021}.

Over the last years, aiming at a more complete theory of fluctuations in graphene, several 
authors have revisited and extended the analysis by considering the effects of 
quantization~\cite{kats_prb_2014, *kats_prb_2014e, coquand_pre_2016, burmistrov_prb_2016} 
and of the coupling between membrane phonons and Dirac electrons~\cite{gazit_prb_2009, 
san-jose_prl_2011, guinea_prb_2014, cea_prb_2020}.
The interaction between flexural and electronic degrees of freedom, in particular, has 
been predicted to generate mechanical instabilities leading to a spontaneous rippling of 
the membrane.

Despite the progress in investigations of the coupled membrane-electron problem, the 
theory of purely mechanical degrees of freedom in a flexible surface subject to both 
thermal and quantum fluctuations is already highly non-trivial.
By a combination of elasticity theory and a one-loop momentum-shell renormalization group, 
Ref.~\cite{kats_prb_2014, *kats_prb_2014e} showed that, for a homogeneous and unstressed 
membrane at absolute zero, mechanical nonlinearities give rise to logarithmic 
renormalizations of the wavevector-dependent bending stiffness and elastic constants, in 
sharp contrast with the much stronger power-law renormalizations induced by thermal 
fluctuations.

In Ref.~\cite{coquand_pre_2016}, the theory of quantum flexible membranes was reanalyzed, 
and extended to the finite temperature case, within the framework of non-perturbative 
renormalization group (NPRG) techniques.
For zero temperature, the weak coupling limit of the NPRG recovers results consistent 
with the momentum-shell predictions of Ref.~\cite{kats_prb_2014, *kats_prb_2014e}.
At non-zero temperature, the NPRG analysis allowed to smoothly interpolate a crossover 
between a short wavelength region of zero-point character, and a long-wavelength region, 
determined by thermal fluctuations.
In more detail, the results of Ref.~\cite{coquand_pre_2016} predicted a RG flow exhibiting 
a first quantum region in which anharmonicities are marginally irrelevant, followed, after 
a smooth crossover, by a classical region in which nonlinearities are relevant, 
destabilize the weak coupling approximation, and drive the system to the universal 
interacting fixed point describing classical thermally-fluctuating 
membranes~\cite{nelson_statistical, aronovitz_prl_1988, aronovitz_jpf_1989, 
guitter_jpf_1989, kownacki_pre_2009, kosmrlj_prb_2016}.
The corresponding correlation functions, in particular, behave in the long-wavelength 
limit according to the anomalous scaling law characteristic of classical membranes: in 
the limit of vanishing wavevector $\bk \to 0$, the effective bending rigidity 
diverges as $\kappa(\bk) \approx k^{-\eta_{*}}$ and the elastic Lam\'{e} constant vanish 
as $\lambda(\bk), \mu(\bk) \approx k^{2 - 2\eta_{*}}$.
A similar picture was derived in Ref.~\cite{burmistrov_prb_2016}, by combining a one-loop 
RG with a physical approximation: the replacement of the full anharmonic free energy with 
a Bose-Einstein function with renormalized phonon dispersions.
Other field-theoretical analyses on quantum flexible membranes, such as the expansion for 
large embedding-space dimension and a generalization of the classical self-consistent 
screening approximation, were developed in Ref.~\cite{guinea_prb_2014}, as a part of a 
wider analysis including the coupling between phonons and Dirac electrons.
We finally note that, by different approaches, Refs.~\cite{amorim_prb_2014, 
bondarev_prb_2018}, have predicted a dynamical behavior qualitatively in contrast with 
Refs.~\cite{kats_prb_2014, kats_prb_2014e, coquand_pre_2016, burmistrov_prb_2016}: that 
flexural phonon modes acquire a non-zero sound velocity and a linear dispersion relation 
$\omega(\bk) \propto |\bk|$.

In parallel with analytical approaches, fluctuations of a quantum graphene sheet have also 
been studied by numerical path-integral simulations based on realistic empirical 
potentials for interatomic interactions in carbon (see, for example, 
Ref.~\cite{hasik_prb_2018, herrero_jcp_2016}).

The objective of this work is to analyze the anharmonic effects in quantum membranes by 
systematic perturbative renormalization group methods.
By a power-counting analysis, we construct an effective renormalizable model which we 
expect to capture the dominant singularities of physical quantities in the limit of low 
energies, momenta, temperatures and tensions.
At $T = 0$, the model is renormalizable in the sense of power counting, although it 
exhibits anisotropic scaling between space and time, in analogy with other theories with 
"weighted power counting"~\cite{zinn-justin_qft, brezin_prb_1976, guitter_jpf_1990, 
anselmi_prd_2007, anselmi_jhep_2008, arav_jhep_2019}.
The corresponding RG equations recover in a systematic framework the earlier results 
derived in Refs.~\cite{kats_prb_2014, kats_prb_2014e, coquand_pre_2016, 
burmistrov_prb_2016}.
We also note that the model is mathematically equivalent to a theory of the decoupled 
lamellar phase of a three-dimensional stack of classical crystalline 
membranes analyzed in Ref.~\cite{guitter_jpf_1989}.

For finite temperatures, we use techniques well-known in the theory of finite size 
scaling and other finite-temperature field theories~\cite{zinn-justin_qft, 
zinn-justin_arxiv_2000}.
In particular, we use the general property that ultraviolet divergences are 
temperature-independent and can be renormalized by $T$-independent 
counterterms~\cite{zinn-justin_qft, zinn-justin_arxiv_2000} to derive scaling laws for 
various thermodynamic quantities: the thermal expansion coefficient $\alphar$, the entropy 
$\Sr$, and the specific heat $\Cr$.
By combining the scaling relations with an analysis of thermal fluctuations, we recover 
the result that, for a membrane subject to zero external tension, $\alphar$ is negative 
and tends to zero in the limit $T \to 0$ as a logarithmic function of 
$T$, as predicted in Ref.~\cite{burmistrov_prb_2016}.

The effective model and the method used to derive scaling equations are intrinsically 
focused on the behavior of thermodynamic quantities in the limit of small temperatures.
Therefore, the theory developed here cannot capture the detailed $T$ dependence of the 
thermal expansion coefficient at moderately high temperatures.
In particular, the question whether $\alphar$ changes sign at a certain 
temperature~\cite{zakharchenko_prl_2009, jiang_jpcm_2015, herrero_jcp_2016, 
bondarev_prb_2018}, is beyond the scope of this work. 
We note, however, that the behavior of out-of-plane fluctuations analyzed here contrasts 
with the prediction of a linear dispersion relation for flexural phonons, which was used 
in the analyses of Refs.~\cite{amorim_prb_2014, bondarev_prb_2018}.

The coefficient of in-plane thermal expansion of graphene has been estimated by a number 
of experimental techniques, both for suspended samples and for samples bound to a 
substrate (see, for example, Refs.~\cite{bao_nature_2009, singh_nanot_2010, 
lopez-polin_nms_2021, yoon_nano-lett_2011, feng_small_2021} and references therein).
Experimental results indicate usually a negative thermal expansion at not too large 
temperatures, although a positive expansion has been identified in 
Ref.~\cite{jean_prb_2013} in the case of graphene on a Ir(111) substrate down to liquid 
helium temperatures.

It would be interesting to test experimentally the prediction of that $\alphar$ is nearly 
temperature-independent (the logarithmic functions of $T$ change very slowly over broad 
temperature scales).
This prediction applies only to membranes without a supporting substrate and without 
stress.
For a nonzero applied tension, the low-temperature behavior of the thermal expansion 
coefficient was predicted to vanish in a faster way as $T \to 0$ in 
Ref.~\cite{burmistrov_prb_2016}.

\section{Model}

To study fluctuations of a quantum membrane, we analyze throughout this work an effective 
low-energy model defined by the path integral 
\begin{equation}\label{Z}
{\cal Z} = \int [{\rm d}\bh(\bx, \tau) {\rm d}u_{\alpha}(\bx, \tau)] {\rm e}^{-S/\hbar}
\end{equation}
and the imaginary-time action
\begin{equation}\label{action0}
\begin{split}
S&[\bh(\bx, \tau), u_{\alpha}(\bx, \tau)] = \int_{0}^{\hbar/(k_{\rm B} \Tr)} {\rm d}\tau 
\int {\rm d}^{2}x \bigg\{\frac{\rhor}{2} \dot{\bh}^{2} \\
& + \frac{\kappar}{2} (\pa^{2}\bh)^{2} + \frac{\lambdar}{2} (u_{\alpha \alpha})^{2} + 
\mur u_{\alpha \beta} u_{\alpha \beta} -\sigmar \pa_{\alpha}u_{\alpha}\bigg\}~.
\end{split}
\end{equation}
The degrees of freedom $u_{\alpha}(\bx, \tau)$ and $\bh(\bx, \tau)$ represent, 
respectively, in-plane and out-of-plane displacements of the mass points in the layer.
The second line of the action represents the standard elastic 
energy~\cite{nelson_statistical, nelson_jpf_1987, shankar_pre_2021} 
of a medium with bending rigidity $\kappar$ and Lam\'{e} coefficients $\lambdar$ and 
$\mur$, and is defined in terms of the strain tensor $u_{\alpha \beta} = 
(\pa_{\alpha}u_{\beta} + \pa_{\beta} u_{\alpha} + \pa_{\alpha}\bh \cdot 
\pa_{\beta}\bh)/2$.
The term $-\sigmar \pa_{\alpha}u_{\alpha}$ describes an externally applied isotropic 
in-plane tension~\cite{shankar_pre_2021}.
A positive $\sigmar > 0$ drives a stretching of the membrane, while $\sigmar < 0$ 
corresponds to a compressive stress, which tends to buckle the system out of plane.
The first term in the action, $\rhor \dot{\bh}^{2}/2$, describes instead the kinetic 
energy of out-of-plane fluctuations, and is proportional to the areal mass density 
$\rhor$ 
and to the square of the out-of-plane velocity $\dot{\bh} = \pa\bh(\bx, \tau)/\pa \tau$.
Although physically $\bh$ is a scalar quantity, we consider in general $\bh$ to be a 
vector with $d_{c}$ components ($d_{c}=1$ for physical membranes embedded in 
three-dimensional space).

To regularize ultraviolet divergences we implicitly assume a large-momentum 
cutoff $\Lambda$ of the order of the inverse lattice spacing.

\subsection{Rescaled units}\label{sec:rescaled-units}

By the change of variables $\tau \to (\rhor/\kappar)^{1/2} \tau$, $\bh \to 
\hbar^{1/2}/(\rhor \kappar)^{1/4} \bh$, $u_{\alpha} \to \hbar/(\rhor \kappar)^{1/2} 
u_{\alpha}$, the reduced action $\cS = S/\hbar$ can be recast as
\begin{equation}\label{action0-rescaled}
\begin{split}
\cS &= \int_{0}^{1/T} {\rm d}\tau \int {\rm d}^{2}x \bigg\{\frac{\dot{\bh}^{2}}{2} + 
\frac{1}{2}(\pa^{2}\bh)^{2} + \frac{\lambda}{2} (u_{\alpha \alpha})^{2}\\
& + \mu u_{\alpha \beta} u_{\alpha \beta} - \sigma \pa_{\alpha} u_{\alpha}\bigg\}~,
\end{split}
\end{equation}
where $u_{\alpha \beta} = (\pa_{\alpha}u_{\beta} + \pa_{\beta} u_{\alpha} + 
\pa_{\alpha}\bh \cdot \pa_{\beta}\bh)/2$, $\lambda = \hbar \lambdar/(\rhor 
\kappar^{3})^{1/2}$, $\mu = \hbar \mur/(\rhor \kappar^{3})^{1/2}$, $T = 
(\rhor/\kappar)^{1/2} k_{\rm B}\Tr/\hbar$, and $\sigma = \sigmar/\kappar$.
After these rescalings, all quantities have a dimension in units of wavevector.
The elastic parameters $\lambda$ and $\mu$, which play the role of coupling constants, 
are dimensionless, while the temperature $T$ and the tension $\sigma$ have the dimension 
of a wavevector squared.

Throughout the rest of this paper, we always use rescaled units, unless explicitly 
mentioned.
Quantities in standard units of measurements are marked with tilde symbols.
The in-plane strain $\tilde{v}$, the Gibbs free energy per unit area $\tilde{\Gr} = 
-k_{\rm B}\Tr A^{-1}\ln {\cal Z}$, the entropy density $\Sr = -\pa \tilde{\Gr}/\pa \Tr$, 
the specific heat $\tilde{C} = \Tr {\rm d}\Sr/{\rm d}\Tr$,  and the thermal expansion 
coefficient $\tilde{\alpha} = 2 {\rm d}\tilde{v}/{\rm d}\Tr$, expressed in conventional 
units, are related to the corresponding rescaled quantities $v$, $\Gr = -T A^{-1}\ln 
{\cal Z}$, $S = -\pa \Gr/\pa T$, $C = T {\rm d}S/{\rm d}T$, $\alpha = 2 {\rm d}v/{\rm 
d}T$ as
\begin{equation}
\begin{split}
\tilde{v} & = \frac{\hbar v}{(\rhor\kappar)^{1/2}} ~, \qquad \tilde{\Gr} = \frac{\hbar 
\kappar^{1/2}}{\rhor^{1/2}} \Gr~, \qquad \Sr = k_{\rm B}S\\
\Cr & = k_{\rm B} C~, \qquad \alphar = k_{\rm B} \alpha/\kappar~.
\end{split}
\end{equation}

\subsection{Derivation of the effective model}

The effective action~\eqref{action0} can be derived from a more complete theory by a 
power counting argument.
Focusing on the case of a vanishing external tension $\sigma = 0$, a more complete model, 
which includes the kinetic energy of in-plane modes is given by the manifestly 
rotationally-invariant action~\cite{burmistrov_prb_2016, burmistrov_aop_2018, 
coquand_pre_2016, saykin_aop_2020}
\begin{equation}\label{action1}
\begin{split}
S[\br(\bx, \tau)]  &= \int_{0}^{\hbar/(k_{\rm B}\Tr)} {\rm d}\tau \int {\rm d}^{2}x 
~\bigg\{\frac{\rhor}{2} \dot{\br}^{2} + \frac{\kappar}{2}(\pa^{2}\br)^{2} \\
& + \frac{\lambdar}{2} (U_{\alpha \alpha})^{2} + \mur U_{\alpha \beta} 
U_{\alpha \beta}\bigg\}~,
\end{split}
\end{equation}
where $\br \in \mathbb{R}^{d}$ denotes fluctuating coordinates in the $d$-dimensional 
ambient space and $U_{\alpha \beta} = (\pa_{\alpha}\br \cdot \pa_{\beta}\br - 
\delta_{\alpha \beta})/2$.
This fully rotationally-invariant theory can be analyzed by parametrizing $\br(\bx, \tau) 
= (\xi \bx + \mathbf{u}(\bx, \tau), \bh(\bx, \tau))$, where $\mathbf{u}$ and $\bh$ 
are in-plane and out-of-plane displacement fields, while $\xi$ encodes the tendency of 
the membrane to shrink due to fluctuations~\cite{guitter_jpf_1989, burmistrov_prb_2016, 
coquand_pre_2016, burmistrov_aop_2018, shankar_pre_2021}.
At zero temperature, a loop expansion (formally an expansion in powers of 
$\hbar$~\cite{zinn-justin_qft}) can be given by calculating order by order correlation 
functions and $\xi =  1 + c_{1} \hbar + c_{2} \hbar^{2} + ..$.
The non-interacting propagators of in-plane and out-of-plane modes, defining the basic 
elements in the corresponding diagrammating expansion, are, respectively
\begin{equation} \label{propagator0}
\begin{split}
\bar{D}^{(0)}_{\alpha \beta}(\omega, \bk) & = \frac{\hbar k_{\alpha} k_{\beta}}{(\rhor 
\omega^{2} + (\lambdar + 2 \mur)|\bk|^{2} + \kappar|\bk|^{4})k^{2}}\\
& + \frac{\hbar (k^{2} \delta_{\alpha \beta} - k_{\alpha} k_{\beta})}{(\rhor \omega^{2} + 
\mur |\bk|^{2} + \kappar |\bk|^{4})k^{2}}~,\\
\bar{G}^{(0)}_{ij}(\omega, \bk) & = \frac{\hbar \delta_{ij}}{\rhor \omega^{2} + \kappar 
|\bk|^{4}}~.\\
\end{split}
\end{equation}
For small $\bk$, $\bar{D}^{(0)}_{\alpha \beta}(\omega, \bk)$ has a pole for $\omega \sim 
|\bk|$ reflecting the linear dispersion of acoustic phonons while $\bar{G}^{(0)}_{\alpha 
\beta}(\omega, \bk)$ has a pole for $\omega \sim k^{2}$, corresponding to the ultrasoft 
dispersion of flexural fluctuations at zero external tension.
Due to the softer infrared behavior of flexural phonons, we can assume that poles of 
$\bar{G}^{(0)}$ generate the leading singularities at long wavelengths.
In the region $\omega \sim k^{2}$, interactions can be analyzed within power counting by 
assigning dimension $[\bx]=-1$ to the spacial coordinates and $[\tau] = -z = -2$ to the 
time coordinate~\cite{coquand_pre_2016}
(see Refs.~\cite{brezin_prb_1976, guitter_jpf_1990, anselmi_prd_2007, anselmi_jhep_2008, 
arav_jhep_2019} for discussions of
of various field theories which lack Lorentz and Euclidean invariance and which exhibit 
"weigthed power counting", with different weigth for space and time coordinates).
The behavior of propagators for $\bk \to 0$, $\omega \to 0$, $\omega \sim k^{2}$ 
implies that the canonical dimensions of fields are, respectively, $[\bh] = (2 + z - 
4)/2 = 0$ and $[u_{\alpha}] = (2 + z - 2)/2 = 1$.
An analysis of dimensions of operators then shows that the elastic parameters $\lambda$ 
and $\mu$ are marginal, whereas the term $\kappar (\pa^{2}u_{\alpha})^{2}/2$, the 
nonlinear contribution $\pa_{\alpha}u_{\gamma}\pa_{\beta}u_{\gamma}$ to the strain 
tensor $U_{\alpha \beta} = ((\xi^{2}-1)\delta_{\alpha \beta} + \xi \pa_{\alpha} u_{\beta} 
+ \xi \pa_{\beta} u_{\alpha} + \pa_{\alpha}\bh \cdot \pa_{\beta} \bh + \pa_{\alpha} 
u_{\gamma}\pa_{\beta} u_{\gamma})/2$, 
and the in-plane kinetic energy $\rhor \dot{u}_{\alpha}^{2}/2$ are all irrelevant in the 
sense of power counting.
By dropping all power-counting irrelevant interactions, we arrive, after a change of 
variables $u_{\alpha} \to u_{\alpha}/\xi - (\xi^{2}-1) x_{\alpha}/(2\xi)$, to the 
effective model~\eqref{action0} with $\sigma = 0$.

Clearly, the effective model cannot describe the dynamics of in-plane phonons, which 
occurs at scales $\omega \sim |\bk|$.
Power counting indicates however that it should capture in an exact way the leading 
singularities at long wavelengths of static correlation functions (diagrams with all 
external frequencies $\omega = 0$), and more generally, singularities of diagrams with 
external legs in the region $\omega \sim k^{2}$~\cite{Note1}, relevant for the behavior of 
flexural phonons.

For simplicity, we will use the effective model~\eqref{action0} also to calculate 
thermodynamical properties of the membrane, such as the entropy and the average projected 
area, at finite temperature $T$ and nonzero tension $\sigma$.
We expect that the theory describes the leading singular behavior of thermodynamic 
quantities for small $T$ and $\sigma$~\cite{Note2}.

As a further remark, we note that, the theory~\eqref{action0} describes only the 
contribution to thermodynamic quantities of membrane-type fluctuations.
In crystal lattices the thermal expansion coefficient and other thermodynamic quantities 
receive additional contributions from the temperature dependence of the interatomic bond 
length.
We expect that these effects are suppressed at small $T$ and do not contribute to the 
dominant low-temperature singularities.
Within a quasi-harmonic theory, the Gr\"{u}neisen parameters associated with optical and 
acoustic in-plane phonons can be assumed to remain finite for $T \to 0$ and $\sigma \to 
0$~\cite{katsnelson_graphene}.
Thus we can estimate that the effects of a temperature dependent bond length vanishes at 
low temperatures proportionally to the specific heat of these phonon 
branches~\cite{katsnelson_graphene}.
By contrast, the fluctuation modes considered here generate infrared singularities, and, 
as shown in Ref.~\cite{burmistrov_prb_2016} and discussed below, produce a contribution to 
the thermal expansion which remains almost constant as $T \to 0$.

\subsection{Symmetries}
In full analogy with the theory of classical membranes, it can be checked that, when  
$\sigma = 0$, $\cS$ is invariant under the "linearized 
rotations"~\cite{aronovitz_prl_1988, guitter_jpf_1989, guitter_jpf_1990}
\begin{equation}\label{rotation}
\begin{split}
\bh & \to \bh + \mathbf{A}_{\alpha} x_{\alpha}~,\\
u_{\alpha} & \to u_{\alpha} - (\mathbf{A}_{\alpha} \cdot \bh) - \frac{1}{2} 
(\mathbf{A}_{\alpha}\cdot \mathbf{A}_{\beta}) x_{\beta}~,
\end{split}
\end{equation}
where $\mathbf{A}_{\alpha}$ is any fixed vector.
This symmetry represents a linearized form of the original SO($d$) invariance of the full 
theory, and reflects the fact that the layer is located in an isotropic ambient space 
(without external forces and with no externally-imposed in-plane tension).
The associated Ward identities~\cite{aronovitz_prl_1988, guitter_jpf_1989, 
guitter_jpf_1990} play a crucial role in the dynamics and the renormalization of the 
model.

As a remark, we note that the linearized invariance~\eqref{rotation} only emerges when 
all irrelevant terms are dropped from the action at the same time.
If, instead, we had neglected the nonlinear contribution 
$\pa_{\alpha}u_{\gamma}\pa_{\beta}u_{\gamma}/2$ to the strain tensor but we had kept the 
kinetic energy of in-plane phonons $\rhor \dot{u}_{\alpha}^{2}/2$, we would have arrived 
at a theory which lacks both the full rotational SO$(d)$ symmetry and the linearized, 
effective rotational symmetry~\eqref{rotation}.
In this case, renormalization would generate generic anisotropic interactions, including 
anisotropies which are relevant in the sense of power counting.
This would then result in an artificial modification of the qualitative behavior of 
fluctuations.
Although the crucial role of symmetries has been appreciated, several approaches in the 
earlier literature used actions or approximations which, in some steps of derivations, 
violate both the exact and the linearized SO$(d)$ symmetries.

In particular, we note that the prediction of a contribution $\Sigma(0,\bk) \propto k^{2}$ 
to the self-energy of flexural phonons at zero frequency, derived in 
Ref.~\cite{amorim_prb_2014}, started from an action in which the nonlinear contribution to 
the strain tensor was neglected but the kinetic energy of in-plane phonons was retained.
An explicit calculation using a full rotationally-invariant 
action~\cite{burmistrov_prb_2016} showed instead that self-energy corrections 
proportional to $k^{2}$ vanish in absence of external stress, consistently with the Ward 
identities~\cite{burmistrov_aop_2018}.
The emergence of the linearized symmetry~\eqref{rotation} ensures that the cancellation 
of terms proportional to $k^{2}$ is consistently captured by the effective 
model~\eqref{action0-rescaled}, as we verify below (see 
Sec.~\ref{sec:two-point-correlations}).

\subsection{Analogy with a model of lamellar phases}
After identification of the imaginary time $\tau$ with an additional space dimension $z$, 
the quantum action $\cS$ turns out to be almost identical to the effective Hamiltonian 
\begin{equation}\label{H-lamellar}
\begin{split}
H & = \int {\rm d}z \int {\rm d}^{2}x 
\bigg\{\frac{B_{0}}{2}(\pa_{z} u_{z})^{2}  + \frac{K_{0}}{2}(\pa_{\perp}^{2}u_{z})^{2} \\
& + \frac{\mu_{0}^{\perp \perp} }{4} (\pa_{\alpha}u_{\beta} + 
\pa_{\beta}u_{\alpha} + \pa_{\alpha}u_{z} \pa_{\beta}u_{z})^{2}\\
& + \frac{\lambda_{0}^{\perp \perp}}{8} (2 \pa_{\alpha}u_{\alpha} + \pa_{\alpha}u_{z} 
\pa_{\alpha}u_{z})^{2}+\\
& + \frac{1}{2}\lambda^{\perp z}_{0} (\pa_{z} u_{z})(2 \pa_{\alpha} u_{\alpha} + 
\pa_{\alpha}u_{z} \pa_{\alpha}u_{z})\bigg\}~,
\end{split}
\end{equation}
which was analyzed by Guitter~\cite{guitter_jpf_1990} as a model for a three-dimensional 
shearless stack of \emph{classical} crystalline membranes.
The identity between the two theories only emerges after irrelevant interactions are 
neglected in both models and under the assumption $\lambda^{\perp z}_{0} =0$.

The theory of shearless stacks of membranes has been a subject of debate and some 
authors~\cite{hatwalne_arxiv_2000, hatwalne_prl_1993} have proposed models which differ 
from Eq.~\eqref{H-lamellar} and thus contrast with the results of 
Ref.~\cite{guitter_jpf_1990}.
Establishing a detailed relation between lamellar phases and quantum membranes is beyond 
the scope of our work.
We will verify, however, that the RG equations for the quantum membrane action recovers 
long-wavelength singularities identical to those predicted in 
Ref.~\cite{guitter_jpf_1990}.

\section{Integration over in-plane modes}
\label{sec:integration-in-plane}

Since $\cS$ is quadratic in $u_{\alpha}$, the in-plane modes can be integrated out 
explicitly.
To integrate out $u_{\alpha}$ it is essential to separate the strain tensor $u_{\alpha 
\beta}$ into uniform modes (with zero spacial momentum $\bk = 0$) and non-uniform 
components (with spacial Fourier components $\bk \neq 0$).
Integration over in-plane phonon modes $u_{\alpha}^{\bk \neq 0}(\bx, \tau)$ with $\bk 
\neq 0$ gives rise to an effective four-point vertex~\cite{guinea_prb_2014}
\begin{equation} \label{vertex-GCI}
\cS^{\bk \neq 0}_{\rm int} = \frac{Y}{8}\int_{0}^{1/T} {\rm d}\tau \sum_{\bk \neq 0} 
P^{T}_{\alpha \beta}(\bk) P^{T}_{\gamma \delta}(\bk) f_{\alpha \beta}(\bk, \tau) 
f_{\gamma \delta}(-\bk, \tau)~,
\end{equation}
where $f_{\alpha \beta}(\bk, \tau)$ is the spacial Fourier transform of the 
composite field $f_{\alpha \beta}(\bx, \tau) = (\pa_{\alpha}\bh (\bx, \tau) \cdot 
\pa_{\beta}\bh (\bx, \tau))$, $P^{T}_{\alpha \beta}(\bk) = \delta_{\alpha \beta} - 
k_{\alpha}k_{\beta}/k^{2}$ is the projector transversal to the momentum transfer $\bk$, 
and $Y = 4 \mu (\lambda + \mu)/(\lambda + 2 \mu)$ is the (dimensionless) Young 
modulus.
The interaction~\eqref{vertex-GCI} represents physically an instantaneous long-range 
coupling between local Gaussian curvatures in the membrane, and is a direct quantum 
generalization of the usual effective interaction which emerges in classical 
theories~\cite{nelson_statistical, nelson_jpf_1987, aronovitz_prl_1988, 
le-doussal_aop_2018, burmistrov_aop_2018, shankar_pre_2021}.

The analysis of zero modes differs depending on the ensemble 
considered (see Ref.~\cite{shankar_pre_2021} for an analysis of isometric and 
isotensional ensemble in the theory of classical membranes).
Here, we find it convenient to use a fixed-stress, or "isotensional" 
ensemble~\cite{shankar_pre_2021}, in which the external in-plane stress $\sigma$ is kept 
fixed and the projected area is allowed to fluctuate.
In this setting, we parametrize $u_{\alpha}(\bx, \tau) = v_{\alpha \beta}x_{\beta} + 
u^{\bk \neq 0}_{\alpha}(\bx, \tau)$ and integrate over all values of both 
$u^{\bk \neq 0}_{\alpha}(\bx, \tau)$ and $v_{\alpha \beta}$.
After integration, we are lead to a contribution to the effective action
\begin{equation}
\int_{0}^{1/T} {\rm d}\tau \int{\rm d}^{2}x \bigg[ \frac{\sigma}{2}(\pa_{\alpha}\bh 
\cdot \pa_{\alpha}\bh) - \frac{\sigma^{2}}{2(\lambda + \mu)}\bigg] + \cS^{\bk = 
0}_{\rm int}~,
\end{equation}
where
\begin{equation}\label{zero-mode-coupling}
\begin{split}
\cS^{\bk=0}_{\rm int} & = A \int_{0}^{1/T} {\rm d}\tau \bigg[\frac{\lambda}{8} 
(f^{0}_{\alpha \alpha} (\tau)- \bar{f}^{0}_{\alpha \alpha})^{2} \\
& + \frac{\mu}{4} (f^{0}_{\alpha \beta}(\tau) - \bar{f}_{\alpha 
\beta}^{0})(f^{0}_{\alpha \beta}(\tau) - \bar{f}_{\alpha \beta}^{0}) \bigg]~,
\end{split}
\end{equation}
$f^{0}_{\alpha \beta}(\tau) = A^{-1} \int {\rm d}^{2}x (\pa_{\alpha}\bh(\bx, \tau) \cdot 
\pa_{\beta}\bh(\bx, \tau))$, and $\bar{f}^{0}_{\alpha \beta} = T \int_{0}^{1/T} {\rm 
d}\tau f^{0}_{\alpha \beta}(\tau)$.
The average strain of the membrane in this ensemble is $\langle v_{\alpha \beta}\rangle 
= v \delta_{\alpha \beta}$, with $v = \sigma/(2(\lambda + \mu)) - \langle 
\pa_{\alpha}\bh \cdot \pa_{\alpha}\bh\rangle/4$~\cite{gornyi_2dmater_2017, 
burmistrov_aop_2018, shankar_pre_2021}.
It is the sum of a Hookean contribution $\sigma/(2(\lambda + \mu))$, controlled by the 
bulk modulus $B = \lambda + \mu$, and a negative fluctuation term, 
proportional to $\langle\pa_{\alpha}\bh\cdot \pa_{\alpha}\bh \rangle$, which is 
nonvanishing also for $\sigma = 0$, and which represents the tendency of the projected 
in-plane area to contract due to statistical fluctuations of the layer in the 
out-of-plane direction.

The infinite-range interaction $\cS^{\bk = 0}_{\rm int}$ is scaled by an overall factor 
$A^{-1}$ and, by its definition, it vanishes when the Matsubara-frequency transfer 
between the composite operators $f_{\alpha \beta}^{0}(\tau)$ is zero.
These two facts together imply that, in the thermodynamic limit $A \to \infty$, 
$\cS^{\bk=0}_{\rm int}$ only contributes via diagrams of the type
\begin{equation}\label{zero-mode-diagrams}
\includegraphics[scale=1.1]{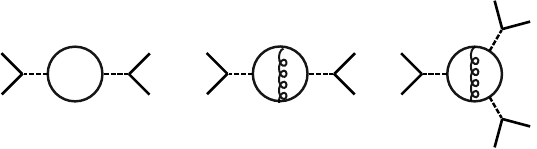}
\end{equation}
which (a) become disconnected when any zero-mode interaction line (represented by dashed 
lines) is cut and (b) have non-zero frequency transfer through all dashed lines. (The 
interaction $\cS^{\bk \neq 0}_{\rm int}$, denoted by wiggly lines can enter, instead, in 
arbitrary topology without suppressing the graphs).
The diagrams~\eqref{zero-mode-diagrams}, however, are only relevant for zero-mode 
correlation functions at finite frequency transfer and never enter as subgraphs of other 
correlation functions.
For subsequent calculations in this work, we can thus safely neglect $\cS^{\bk = 0}_{\rm 
int}$.

As a result, we can thus consider an effective theory for $\bh$ fluctuations of the form:
\begin{equation}
\begin{split}
\cS_{\rm eff} & = \int_{0}^{1/T} {\rm d}\tau \int {\rm d}^{2}x 
\bigg\{\frac{\dot{\bh}^{2}}{2} + \frac{1}{2} (\pa^{2}\bh)^{2} \\
& + \frac{\sigma}{2}(\pa_{\alpha}\bh)^{2} - \frac{\sigma^{2}}{2B}\bigg\} + \cS^{\bk 
\neq 0}_{\rm 
int}~.
\end{split}
\end{equation}
By a Hubbard-Stratonovich decoupling of the long-range interaction~\cite{mauri_npb_2020}, 
the model can be expressed equivalently via the local action
\begin{equation}\label{action-GCI}
\begin{split}
\cS_{\rm eff} & = \int_{0}^{1/T} {\rm d}\tau \int {\rm d}^{2}x  
\bigg\{\frac{\dot{\bh}^{2}}{2} + \frac{1}{2}(\pa^{2}\bh)^{2} + 
\frac{\sigma}{2}(\pa_{\alpha}\bh)^{2} \\
& + \frac{1}{2Y} (\pa^{2}\chi)^{2} + i \chi K - \frac{\sigma^{2}}{2B}\bigg\}~,
\end{split}
\end{equation}
where $\chi(\bx, \tau)$ is a mediator field and $K(\bx, \tau) = (\pa^{2}\bh \cdot 
\pa^{2}\bh - \pa_{\alpha}\pa_{\beta}\bh \cdot \pa_{\alpha}\pa_{\beta}\bh)/2$ is, for 
small fluctuations, the local Gaussian curvature.
The term $-\sigma^{2}/(2B)$ is a constant independent of the fluctuating fields $\bh$, 
$\chi$, and does not contribute to statistical averages.
The only coupling constant in the model is thus the Young modulus $Y$.

By construction, the interaction-mediating field $\chi$ must be considered as a field 
with Fourier components only at nonzero momentum $\bk \neq 0$.
This implies that the tadpole graphs
\begin{equation}
\includegraphics[scale=1]{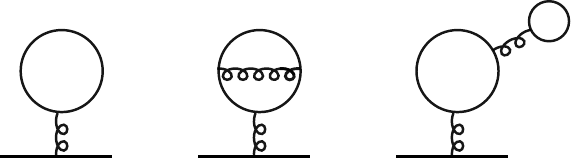}
\end{equation}
must be removed from the perturbative expansion, as in the theory of classical 
membranes~\cite{aronovitz_jpf_1989}.

\section{Renormalization and RG equations at zero temperature} 
\label{sec:renormalization_T=0}

\subsection{RG for correlation functions}

At $T = 0$ the model is infinite in both spacial and temporal dimensions, and its 
renormalization can proceed in analogy with other bulk theories with  
weighted power counting~\cite{zinn-justin_qft, brezin_prb_1976, guitter_jpf_1990, 
anselmi_prd_2007, anselmi_jhep_2008, arav_jhep_2019}.
In the representation~\eqref{action-GCI}, the basic elements defining diagrams in 
perturbation theory are the bare propagators of $\bh$ and $\chi$,
\begin{equation}\label{propagators}
\begin{split}
& \includegraphics[scale=1]{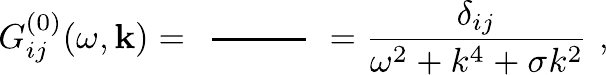}\\
&\includegraphics[scale=1]{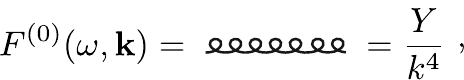}\\
\end{split}
\end{equation}
and the vertex
\begin{equation*}
\includegraphics[scale=1]{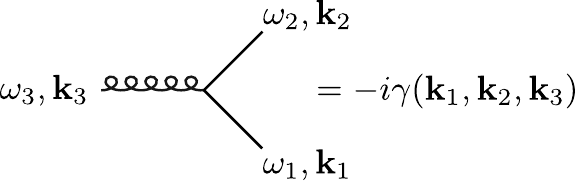}
\end{equation*}
\begin{equation}\label{GCI-vertex}
\begin{split}
\gamma(\bk_{1}, \bk_{2}, \bk_{3}) & = k_{1}^{2} k_{2}^{2} - (\bk_{1}\cdot 
\bk_{2})^{2} = k_{2}^{2} k_{3}^{2} - (\bk_{2}\cdot \bk_{3})^{2} \\
& = k_{3}^{2} k_{1}^{2} - (\bk_{3} \cdot \bk_{1})^{2}~.
\end{split}
\end{equation}
The behavior of Feynman integrals under the rescaling $\bk \to \lambda \bk$, $\omega \to 
\lambda^{z} \omega = \lambda^{2} \omega$ shows that the weighted power-counting 
dimension~\cite{anselmi_prd_2007} of a one-particle irreducible (1PI) diagram with $I$ 
internal lines, $V$ vertices, and $L$ loops is $(2+z) L - 4 I + 4 V = 4 (L - I + V) = 4$, 
independently of the order of perturbation theory.
This ensures that the model is power-counting renormalizable.
A potential danger for renormalizability~\cite{anselmi_jhep_2008} is that the propagator 
$F^{(0)}(\omega, \bk)$, being $\omega$-independent, is not suppressed in the limit 
$\omega \to \infty$ at $\bk$ fixed.
However, this does not create difficulties, because it can be checked that in any 
diagram, all frequency integrals can be performed first and are convergent~\cite{Note3}.

To complete the proof of renormalizability, it would be necessary to derive a 
generalization of the Weinberg theorem~\cite{weinberg_pr_1960, hahn_cmp_1968}, ensuring 
the equivalence between power-counting convergence and true convergence in multiloop 
diagrams.
We will assume that this property remains valid in the model considered in this work.

The ultraviolet divergences of correlation functions can be removed by introducing an 
arbitrary subtraction scale $M$, a renormalized coupling $Y_{\rm R}$, and a renormalized 
action
\begin{equation}\label{S-renormalized}
\begin{split}
\cS_{\rm eff}^{\rm (R)} & = \int {\rm d}\tau \int {\rm d}^{2}x 
\bigg\{\frac{\dot{\bh}^{2}}{2} + \frac{Z}{2} (\pa^{2}\bh)^{2} + \frac{\sigma}{2} 
(\pa_{\alpha}\bh)^{2} \\
& + \frac{1}{2K_{Y}} (\pa^{2}\chi)^{2} + i \chi K - \frac{\sigma^{2}}{2B}\bigg\}~,
\end{split}
\end{equation}
equipped with two logarithmically divergent counterterms $Z(Y_{\rm R}, \Lambda/M)$ and 
$K_{Y}(Y_{\rm R}, \Lambda/M)$.
This particularly simple form, with only two independent divergences, follows from the 
fact that the terms $\dot{\bh}^{2}/2$, $\sigma (\pa_{\alpha}\bh)^{2}/2$, and the 
interaction $i \chi K$ are not renormalized.
Indeed, due to the structure of the vertex~\eqref{GCI-vertex}, it is possible to 
factorize, from any 1PI diagram, two powers of the spacial momentum of each external leg.
Therefore, the perturbative corrections to the self-energy of flexural fields cannot 
generate divergences proportional to $\omega^{2}$ or to $k^{2}$, but only proportional to 
$k^{4}$, which contribute to the renormalization of $Z$.
The possibility to factorize two powers of each external momentum also implies that loop 
corrections to the three-field vertex are superficially convergent, and thus the 
interaction $i \chi K$ does not require an independent counterterm.
An identical mechanism occurs in the $\varepsilon$-expansion of classical membranes in 
dimension $D = 4 -\varepsilon$~\cite{le-doussal_aop_2018, mauri_npb_2020}.
In principle, the one-point function $\langle \pa^{2}\chi\rangle$ constitutes a further 
independent divergence, but since $\chi$ is a field with components only at nonzero 
momentum $\bk \neq 0$, this divergence is unphysical and has no effect on correlation 
functions.

Eq.~\eqref{S-renormalized} implies the following relations between bare and renormalized 
quantities
\begin{equation}
\begin{split}
&\cS_{\rm eff}[\bh, \chi, Y, \sigma]  = \cS_{\rm eff}^{({\rm R})}[\bh_{\rm R}, 
\chi_{\rm R}, Y_{\rm R}, \sigma_{\rm R}] + \text{constant}~,\\
&\bh(\bx, \tau)  = Z^{1/4}\bh_{\rm R}(\bx, \tau_{\rm R})~, \qquad 
\chi(\bx, \tau) = Z^{-1} \chi_{\rm R}(\bx, \tau_{\rm R})~,\\
& \tau  = Z^{1/2}\tau_{\rm R}~, \qquad \sigma = Z^{-1}\sigma_{\rm R}~, \qquad Y = 
Z^{-3/2}K_{Y}
\end{split}
\end{equation}
and, according to standard techniques~\cite{zinn-justin_qft}, the following RG 
equations for 1PI correlation functions in momentum space
\begin{equation}\label{RG-GCI}
\begin{split}
&\bigg[ \Lambda\frac{\pa}{\pa \Lambda} + \beta(Y) \frac{\pa}{\pa Y} - \frac{1}{2}(n - 
4 \ell + 2) \eta + \eta \sum_{i = 1}^{n} \omega_{i} \frac{\pa}{\pa \omega_{i}} \\
& + \eta \sum_{j = 1}^{\ell} \omega'_{j} \frac{\pa}{\pa \omega'_{j}} + 2 \eta
\sigma\frac{\pa}{\pa \sigma}\bigg] \Gamma^{(n, \ell)}_{i_{1} .. i_{n}}(\omega_{i}, 
\bk_{i}; \omega'_{j}, \bk'_{j}) = 0~.
\end{split}
\end{equation}
In Eq.~\eqref{RG-GCI}, $\Lambda$ is the microscopic ultraviolet momentum cutoff and 
$\Gamma^{(n, \ell)}$ denotes the bare (unrenormalized) 1PI correlation function with $n$ 
external $\bh$ legs and $\ell$ external $\chi$ legs.
The RG flow function $\beta  = \Lambda \pa Y/\pa \Lambda$ and the anomalous 
dimension $\eta = - \frac{1}{2} \Lambda \pa \ln Z/\pa \Lambda$ depend only on the bare 
dimensionless coupling $Y$.

By an explicit computation of the one-loop divergences in the self-energies of $\chi$ and 
of $\bh$ for $\sigma=0$ we find~\cite{kats_prb_2014, guinea_prb_2014, 
burmistrov_prb_2016}
\begin{equation}
\begin{split}
&\includegraphics[scale=1]{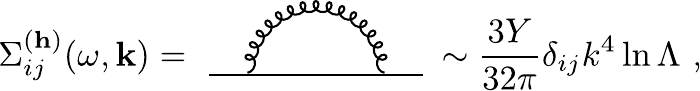}\\
&\includegraphics[scale=1]{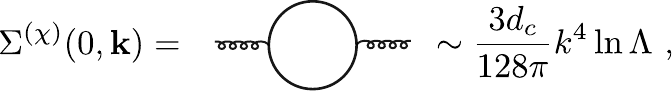}\\
\end{split}
\end{equation}
where $d_{c} = (d-2)$ is the number of components of the $\bh$ field.
Applying the RG equations to $G_{ij}^{-1}(\omega, \bk) = \Gamma^{(2, 0)}_{ij}(\omega, 
\bk) = \omega^{2} + k^{4} + \sigma k^{2} + \Sigma^{(\bh)}(\omega, \bk)$ and $F^{-1}(0, 
\bk) = \Gamma^{(0, 2)}(0, \bk) = k^{4}/Y + \Sigma^{(\chi)}(0, \bk)$ shows that, at 
leading order,
\begin{equation}\label{1L-RG}
\beta(Y) = 3 \eta(Y) Y + \frac{3 d_{c}}{128 \pi} Y^{2} ~, \qquad \eta(Y) = 
\frac{3}{64 \pi}Y~.
\end{equation}
These RG functions imply that $Y$ is marginally irrelevant: the theory is attracted to 
weak coupling at large length scales.
Eqs.~\eqref{1L-RG} are consistent, in a different scheme, with the earlier perturbative 
results of Refs.~\cite{kats_prb_2014, burmistrov_prb_2016} and also with the 
weak-coupling limit of the nonperturbative RG equations derived in 
Ref.~\cite{coquand_pre_2016}.

\subsection{Gibbs free energy}

The Gibbs free energy per unit area at zero temperature, $\Gr_{0} = -A^{-1} \lim_{T \to 
0} (T \ln {\cal Z})$, requires the introduction of additional counterterms to the 
field-independent part of the action.
Since $\sigma$ has power-counting dimension $[\sigma] = 2$, the required counterterms
in the Lagrangian are a polynomial $a_{0} + a_{1} \sigma + a_{2} \sigma^{2}/2$, where 
$a_{0}$ diverges as $\Lambda^{4}$, $a_{1}$ as $\Lambda^{2}$, and $a_{2}$ diverges 
logarithmically.
By working within a massless scheme~\cite{zinn-justin_qft, anselmi_prd_2007}, 
$a_{0}$, 
$a_{1}$, and $a_{2}$ can be chosen to be independent of the tension $\sigma$.

Taking into account these additional renormalizations, the renormalized action reads
\begin{equation}\label{action-GCI-R}
\begin{split}
& \cS^{({\rm R})}_{\rm eff} = \int {\rm d}\tau \int {\rm d}^{2}x 
\bigg\{\frac{\dot{\bh}^{2}}{2} + \frac{Z}{2} (\pa^{2}\bh)^{2} + \frac{\sigma}{2} 
(\pa_{\alpha}\bh)^{2} \\
& + \frac{1}{2 K_{Y}} (\pa^{2} \chi)^{2} + i \chi K + a_{0} + a_{1} \sigma + 
\frac{1}{2}(a_{2}-1/B) \sigma^{2} \bigg\}~.
\end{split}
\end{equation}
To discuss RG equations it is convenient to separate $\Gr_{0} = V_{\rm el} + \Delta 
\Gr_{0}$, where $V_{\rm el} = - \sigma^{2}/(2B)$ is the elastic Hookean contribution 
and $\Delta \Gr_{0}$ is the fluctuation part.
The advantage of this separation is that $\Delta \Gr_{0}$ depends only on the Young 
modulus $Y$ and not on the bulk modulus $B$.

The fluctuation free energy $\Delta \Gr^{({\rm R})}_{0}$, calculated using the 
action~\eqref{action-GCI-R}, is finite for $\Lambda \to \infty$ at fixed $Y_{\rm R}$ 
and $M$.
The physical free energy $\Delta \Gr_{0}$, computed from the bare 
action~\eqref{action-GCI} is related to $\Delta \Gr^{({\rm R})}_{0}$ by
\begin{equation}\label{G-renormalization-relation}
\begin{split}
Z^{1/2}\Delta \Gr_{0}(Y, Z^{-1} \sigma, \Lambda) & = \Delta \Gr_{0}^{({\rm 
R})}(Y_{\rm R}, \sigma, M) \\
& - a_{0}' - a_{1} \sigma - \frac{a_{2}}{2} \sigma^{2}~.
\end{split}
\end{equation}
($a_{0}'$ differs from $a_{0}$ because it receives contributions from the path-integral 
measure during the change of variables $\bh \to \bh_{\rm R}$, $\chi \to \chi_{\rm R}$).
The relation~\eqref{G-renormalization-relation} and the finiteness of $\Delta 
\Gr_{0}^{({\rm R})}$ imply an inhomogeneous RG equation for the physical Gibbs free energy
\begin{equation}\label{RG-Gibbs-0}
\begin{split}
\bigg[& \Lambda \frac{\pa}{\pa \Lambda} + \beta(Y) \frac{\pa}{\pa Y} + 2 \eta 
\sigma\frac{\pa}{\pa \sigma} - \eta\bigg] \Delta \Gr_{0}(Y, \sigma, \Lambda) \\
& = b_{0} + b_{1} \sigma  + \frac{b_{2}}{2} \sigma^{2}~.
\end{split}
\end{equation}
The constants $b_{0}$, $b_{1}$, and $b_{2}$ are independent of $\sigma$ in the massless 
scheme and cannot depend on the arbitrary subtraction scale $M$.
Thus they have the form $b_{0} = \bar{b}_{0}(Y) \Lambda^{4}$, $b_{1} = \bar{b}_{1}(Y) 
\Lambda^{2}$, and $b_{2} = \bar{b}_{2}(Y)$.

\section{RG for low-temperature thermodynamic quantities}\label{sec:RG-low-T}

At finite temperatures, the continuum frequency $\omega$ is replaced by discrete 
bosonic Matsubara frequencies $\omega_{n} = 2 \pi T n$.
As a result, even for an infinitesimal $T$, the perturbative expansion at zero 
tension $\sigma = 0$ breaks down due to infrared (IR) divergences.
The IR problems arise from the $\omega_{n} = 0$ component of the flexural propagator 
$G^{(0)}_{ij} = \delta_{ij}/(\omega_{n}^{2} + k^{4})$, which induces singularities when 
integrated over the two-dimensional spacial momenta.
The physical origin of these divergences is the following: in the limit $\bk \to 0$, the 
system behaves as a classical membrane~\cite{coquand_pre_2016}.
For classical thermal fluctuations, anharmonic effects do not induce logarithmic 
corrections but, rather, power-law renormalizations~\cite{nelson_statistical, 
nelson_jpf_1987, aronovitz_prl_1988, aronovitz_jpf_1989, guitter_jpf_1989, 
le-doussal_aop_2018, saykin_aop_2020}.
The dramatic power-law singularities of the classical theory cannot be captured by a 
simple perturbative treatment, but require more detailed solutions, for example within 
the framework of the self-consistent screening approximation~\cite{le-doussal_aop_2018}, 
the non-perturbative RG~\cite{kownacki_pre_2009, coquand_pre_2016}, the large-$d$ 
expansion~\cite{david_epl_1988, saykin_aop_2020}, or the 
$\varepsilon$-expansion~\cite{aronovitz_prl_1988, mauri_npb_2020, coquand_pre_2020, 
metayer_pre_2022, pikelner_arxiv_2021}.

Similar difficulties emerge in finite-size scaling problems and in other 
finite-temperature quantum field theories.
A standard strategy to bypass the problem of IR singularities consists in integrating out 
modes with $n \neq 0$ and in deriving an effective field theory for modes with $n = 0$, 
to be solved by more exact methods~\cite{zinn-justin_qft}.

The same strategy can be applied to the membrane action~\eqref{action-GCI}, with, 
however, a difference compared to the standard case: since the $\chi$ propagator 
$F^{(0)}(\omega, \bk) = Y/k^{4}$ does not depend on the frequency $\omega$, it is 
singular at small $\bk$ not only at $\omega_{n} = 0$ but, in fact, for all Matsubara 
frequencies $\omega_{n} \neq 0$.
As a result, subtracting the modes $\omega_{n} = 0$ does not introduce an IR cutoff to 
Feynman diagrams.
This property is a consequence of the neglection of the kinetic energy of in-plane 
phonons.
The singularity of $F^{(0)}(\omega, \bk)$, however, is neutralized by the factors $k^{2}$ 
attached to the vertices~\eqref{GCI-vertex} and, therefore, the IR finiteness is still 
valid.

We can thus proceed as follows: we separate $\bh(\bx, \tau) = \bh'(\bx, \tau) + 
\bH(\bx)$, where $\bH(\bx) = T \int _{0}^{1/T} \bh(\bx, \tau)$ is the mode with zero 
Matsubara frequency and $\bh'(\bx, \tau)$ the sum of all other modes with $\omega_{n} 
\neq 0$.
We then integrate out $\bh'(\bx, \tau)$ and \emph{all} degrees of freedom of $\chi(\bx, 
\tau)$ (including the $\omega_{n} = 0$ mode of $\chi$).
This integration can be performed perturbatively without encountering IR divergences 
because in all $\bh'$ propagators the finite frequency $\omega_{n} \neq 0$ provides an IR 
cutoff and in all $\chi$ propagators the singularity $Y/k^{4}$ of the propagator is 
compensated by a power $k^{4}$ coming from the vertex~\eqref{GCI-vertex}.
Although the singularity of $F^{(0)}(\omega, \bk)$ does not introduce divergences, it 
still manifests itself in the fact that the effective theory for $\bH(\bx)$ is 
highly non-local.

In order to disentangle modes which generate IR singularities from degrees of freedom 
which generate UV divergences, it is also convenient to separate $\bH(\bx) = 
\bH_{1}(\bx) + \bH_{2}(\bx)$ into a slowly-varying field $\bH_{1}(\bx)$, with momenta 
$|\bk| < \Lambda_{1}$ and a fast field $\bH_{2}(\bx)$ with momenta in the shell 
$\Lambda_{1}<|\bk|<\Lambda$, where $\Lambda_{1}$ is an arbitrary wavevector scale much 
smaller than $\Lambda$.
Integrating out $\bh'$, $\chi$, and $\bH_{2}$ can be done perturbatively and leaves us 
with an effective classical Hamiltonian
\begin{equation}\label{H1}
\begin{split}
 {\cal H}&[\bH_{1}(\bx); Y, \sigma, T, \Lambda, \Lambda_{1}]  \\
& = - T \ln \int [{\rm d}\bh' {\rm d}\bH_{2} {\rm d}\chi] {\rm e}^{-\cS[\bH_{1} + 
\bH_{2} + \bh', \chi]}~.
\end{split}
\end{equation}
involving only slowly-varying long-wavelength modes.

A crucial observation in the theory of finite-size scaling and other finite-temperature 
field theories is that the counterterms which make the theory finite at $T = 0$ will also 
formally remove all ultraviolet divergences from observables at nonzero 
$T$~\cite{zinn-justin_qft, zinn-justin_arxiv_2000}.
It is natural to assume that the same property remains valid for the membrane action.
We can thus conclude that if we started from the action
\begin{equation}\label{action-GCI-R-T}
\begin{split}
& \cS^{({\rm R})}_{\rm eff} = \int_{0}^{1/T} {\rm d}\tau \int {\rm d}^{2}x 
\bigg\{\frac{\dot{\bh}^{2}}{2} + \frac{Z}{2} (\pa^{2}\bh)^{2} + \frac{\sigma}{2} 
(\pa_{\alpha}\bh)^{2} \\
& + \frac{1}{2 K_{Y}} (\pa^{2} \chi)^{2} + i \chi K + a_{0} + a_{1} \sigma + 
\frac{1}{2}(a_{2}-1/B) \sigma^{2} \bigg\}~,
\end{split}
\end{equation}
equipped with the same zero-temperature counterterms $Z$, $K_{Y}$, $a_{0}$, $a_{1}$, and 
$a_{2}$ which appear in Eq.~\eqref{action-GCI-R}, after a perturbative integration over 
$\bh'$, $\chi$, and $\bH_{2}$, and a final non-perturbative integration over $\bH_{1}$ 
we would arrive at a renormalized Gibbs free energy per unit area $\Gr^{({\rm R})}= 
-A^{-1} T \ln {\cal Z}$ which remains finite for $\Lambda \to \infty$.

After separation of $\Gr = V_{\rm el} + \Delta \Gr$ into the Hookean part 
$-\sigma^{2}/(2B)$ and the fluctuation part $\Delta \Gr$, the physical fluctuation 
free energy $\Delta \Gr$, computed from the bare action~\eqref{action-GCI} is related to 
the renormalized $\Delta \Gr^{({\rm R})}$ by the equation
\begin{equation}\label{G-finite-T}
\begin{split}
\Delta \Gr^{({\rm R})}(Y_{\rm R}, \sigma, T, M)  & = Z^{1/2}\Delta \Gr(Y, Z^{-1} 
\sigma, Z^{-1/2} T, \Lambda) \\
& + a_{0}' + a_{1} \sigma + \frac{a_{2}}{2} \sigma^{2}~.
\end{split}
\end{equation}
From Eq.~\eqref{G-finite-T} follows an inhomogeneous RG equation for the bare Gibbs free 
energy:
\begin{equation}\label{RG-Gibbs-T}
\begin{split}
\bigg[& \Lambda \frac{\pa}{\pa \Lambda} + \beta(Y) \frac{\pa}{\pa Y} + 2 \eta \sigma 
\frac{\pa}{\pa \sigma} \\
& + \eta T \frac{\pa}{\pa T} - \eta\bigg] \Delta \Gr(Y, \sigma, T, \Lambda) = 
b_{0} + b_{1} \sigma + \frac{b_{2}}{2} \sigma^{2}~.
\end{split}
\end{equation}
In Eq.~\eqref{RG-Gibbs-T}, $\beta(Y)$, $\eta(Y)$, and the coefficients of 
the inhomogeneous part $b_{0} = \bar{b}_{0}(Y)\Lambda^{4}$, $b_{1} = 
\bar{b}_{1}(Y)\Lambda^{2}$, and $b_{2} = \bar{b}_{2}(Y)$ are the same RG coefficients 
which appear in the zero-temperature equation~\eqref{RG-Gibbs-0} and, in particular, are 
temperature-independent.

As a remark, we note that the RG equations discussed above, as in any renormalizable 
theory~\cite{zinn-justin_qft}, keep track of all terms which either diverge or remain 
finite when $\Lambda \to \infty$.
Terms which vanish for large cutoff (for example a correction $\sigma/\Lambda^{2}$) are 
instead neglected.
As a result, relations such as Eq.~\eqref{RG-Gibbs-T} are valid asymptotically when the 
cutoff $\Lambda$ is much larger than other scales in the problem: $\Lambda \gg 
\sigma^{1/2}$, $\Lambda \gg T^{1/2}$.
In standard units of measurement the condition $\Lambda \gg T^{1/2}$ implies that the 
temperature $\Tr$ must be much smaller than the Debye temperature of flexural phonons 
$\Tr_{\rm D} = \hbar \kappar^{1/2} \Lambda^{2}/(\rhor^{1/2} k_{\rm B})$.

\subsection{RG equation for the effective classical Hamiltonian}

More generally, the effective classical Hamiltonian~\eqref{H1} must, by itself, satisfy 
a renormalization group equation
\begin{equation} \label{RG-H1}
\begin{split}
\bigg[& \Lambda\frac{\pa}{\pa \Lambda} + \beta(Y)\frac{\pa}{\pa Y} + 2 \eta 
\sigma\frac{\pa}{\pa \sigma} + \eta T\frac{\pa}{\pa T} \\
& - \frac{\eta}{2} \int {\rm d}^{2}x ~\bH_{1}(\bx) \cdot \frac{\delta}{\delta 
\bH_{1}(\bx)} + \eta \bigg] \Delta {\cal H} \\
& = A \bigg(b_{0} + b_{1}\sigma + \frac{b_{2}}{2} \sigma^{2}\bigg)~,
\end{split}
\end{equation}
where $\Delta {\cal H} = {\cal H} + A \sigma^{2}/(2B)$ is the fluctuation energy, with 
the Hookean contribution $A V_{\rm el} = -A \sigma^{2}/(2B)$ subtracted.
Eq~\eqref{RG-H1} expresses that the cutoff dependence of ${\cal H}$ is entirely carried 
by the zero-temperature counterterms $Z$, $K_{Y}$, $a_{0}$, $a_{1}$, $a_{2}$.

For the direct validity of Eq.~\eqref{RG-H1}, it is essential that all high-energy modes 
are integrated out, as in Eq.~\eqref{H1}.
If, for example, we did not integrate out the large-momentum modes $\bH_{2}(\bx)$ with 
zero Matsubara frequency, we would have moved some of the UV infinities from the 
integrated modes to the degrees of freedom yet to be integrated.
In this case, ${\cal H}$ would have included additional 
counterterms~\cite{zinn-justin_arxiv_2000}.

\section{Results} \label{sec:results}

In this section, we derive explicit consequences of the RG relations for various 
statistical and thermodynamic quantities.

\subsection{Two-point correlation functions at $T = 0$, $\sigma=0$, $\omega = 0$}
\label{sec:two-point-correlations}

The interacting Green functions $G_{ij}(\omega, \bk)$ of the flexural field is the 
inverse of the 1PI function $\Gamma^{(2, 0)}_{ij}(\omega, \bk)$.
For $T = 0$, $\sigma = 0$, and $\omega = 0$, $G_{ij}(0, \bk) = [\Gamma^{(2, 
0)}_{ij}(0, \bk)]^{-1}$ satisfies, as a particular case of Eq.~\eqref{RG-GCI}, the RG 
equation
\begin{equation}\label{RG_G-1}
\Bigg[\Lambda \frac{\pa}{\pa \Lambda} + \beta(Y) \frac{\pa}{\pa Y} - 2 \eta \Bigg] 
G^{-1}_{ij}(0, \bk; Y, \Lambda) = 0~.
\end{equation}
The renormalization group equations can be solved, in an usual way~\cite{zinn-justin_qft}, 
by introducing a running coupling $y(\Lambda')$ and an amplitude renormalization 
$z(\Lambda')$ which, starting from the initial values $y(\Lambda) = Y$, $z(\Lambda) = 
1$, evolve with the floating cutoff scale $\Lambda'$ according to the flow equations
\begin{equation}\label{running-couplings}
\begin{split}
\Lambda'\frac{{\rm d} y(\Lambda')}{{\rm d} \Lambda'} & = \beta(y(\Lambda'))~, \\
\Lambda' \frac{{\rm d} \ln z(\Lambda')}{{\rm d} \Lambda'} & = -2\eta(y(\Lambda'))~.
\end{split}
\end{equation}
The one-loop RG flow gives
\begin{equation} \label{1L-renormalizations}
\begin{split}
y(\Lambda') &= \frac{Y}{1 + \frac{3 (d_{c}+6) Y}{128 \pi}\ln 
\frac{\Lambda}{\Lambda'}}~,\\
z(\Lambda') & = \bigg[1 + \frac{3 (d_{c}+6) Y}{128 
\pi}\ln\frac{\Lambda}{\Lambda'}\bigg]^{\theta} = \lt(\frac{Y}{y(\Lambda')}\rt)^{\theta}~,
\end{split}
\end{equation}
where $\theta = 4/(d_{c}+6)$ is the quantum exponent.

To calculate $G_{ij}(0, \bk)$, we can integrate the RG flow down to a scale $\Lambda' 
\approx k$.
Since $y(\Lambda')$ flows to small values as $\Lambda'$ is reduced (it is marginally 
irrelevant), we can use perturbation theory and take the zero-order approximation 
$G^{-1}_{ij}(0, \bk; y(\Lambda'), \Lambda') = k^{4}$.
The scaling relation~\eqref{RG_G-1} then implies $G_{ij}(0, \bk; Y, \Lambda) \approx 
\delta_{ij}/(z(k) k^{4})$.
By similar arguments, we find that the two point function $F(0, \bk) = \langle \chi(0,
\bk) \chi(0, -\bk)\rangle$ of the auxiliary field $\chi$ scales as $F(0, \bk) \approx 
z^{3/2}(k) y(k)/k^{4}$.

The scaling of $G_{ij}(0, \bk)$ shows that $z(k)$ plays the role of a bending-rigidity 
renormalization and $z^{3/2}(k) y(k)$ the role of an effective screened Young modulus.
Returning to standard units, these results can thus be interpreted as a renormalization 
of the bending rigidity
\begin{equation}\label{kappaR}
\kappar \to \kappar_{\rm r}(k) = \bigg[1 + g_{0} \ln 
\frac{\Lambda}{k}\bigg]^{\theta}\kappar
\end{equation}
and of the elastic Young modulus $\Yr = 4 \mur(\lambdar + \mur)/(\lambdar + 2 \mur)$
\begin{equation}\label{YR}
\tilde{Y} \to \tilde{Y}_{\rm r}(k) = \bigg[1 + g_{0} \ln 
\frac{\Lambda}{k}\bigg]^{3\theta/2-1 } \tilde{Y}
\end{equation}
where
\begin{equation}
g_{0} = \frac{3(d_{c}+6)Y}{128 \pi} = \frac{3(d_{c}+6) \hbar \tilde{Y}}{128 \pi (\rhor 
\kappar^{3})^{1/2}}
\end{equation}
is the "quantum coupling constant"~\cite{burmistrov_prb_2016}.
The bending rigidity gets stiffened by interactions and scales for $k \to 0$ as $[\ln 
(\Lambda/k)]^{\theta}$.
The Young modulus $Y_{\rm r}(k)$, instead, is softened by fluctuations and behaves in 
the long-wavelength limit as $[\ln (\Lambda/k)]^{3\theta/2 - 1} = [\ln 
(\Lambda/k)]^{-d_{c}/(6+d_{c})}$.

The same behavior has been predicted for quantum membranes in 
Refs.~\cite{burmistrov_prb_2016, coquand_pre_2016}.
An identical logarithmic singularity has also been found, for $d_{c}=1$, by 
Ref.~\cite{guitter_jpf_1990} in the context of lamellar stacks of membranes.

The logarithmic renormalizations induced by quantum fluctuations are nonvanishing at all 
momenta and do not exhibit any characteristic crossover scale, due to the simple 
structureless form of the beta function $\beta(Y) \propto Y^{2}$.
This contrasts with the anharmonic renormalizations in classical 
membranes~\cite{nelson_statistical, aronovitz_jpf_1989, le-doussal_aop_2018, 
shankar_pre_2021, burmistrov_aop_2018}, which present a crossover between harmonic 
behavior for $|\bk| \gg q_{G}$ and anomalous power-law scaling for $|\bk| \ll q_{G}$. 
Eqs.~\eqref{kappaR},~\eqref{YR} are valid at all momenta well below to the cutoff scale ( 
for $|\bk| \approx \Lambda$, the continuum approximation breaks down).
Using the explicit expression~\eqref{YR}, we can estimate that the momentum scale scale at 
which the running coupling has reached half of its bare value is approximately $|\bk| 
\approx k_{1} = \Lambda {\rm e}^{-1/g_{0}}$.
The momentum scale $k_{1}$, however, does not mark a special point in the $\bk$-dependence 
of correlation functions.

\subsubsection{Ultrasoft scaling of $G^{-1}(0, \bk)$} \label{sec:ultrasoft}

From the result $G^{-1}_{ij}(0, \bk) \approx k^{4} [\ln(\Lambda/k)]^{\theta}$ it follows, 
in particular, that the ultrasoft behavior $\lim_{\bk \to 0} G^{-1}_{ij}(0, \bk)/k^{2} = 
0$ characteristic of unstressed membranes is preserved by anharmonic effects.
This result is consistent with the general Ward identity $\lim_{\bk \to 0} G^{-1}_{ij}(0, 
\bk)/k^{2} = \sigma$, which is a consequence of rotational 
invariance~\cite{burmistrov_aop_2018} and which, here, can be traced to the linearized 
rotational symmetry~\eqref{rotation} of the effective model~\eqref{action0}.

We note, instead, that this limiting behavior contrasts with the derivations in 
Refs.~\cite{amorim_prb_2014, bondarev_prb_2018}, which proposed that, even in an 
unstressed membrane, flexural phonons exhibit a finite acoustic sound velocity $v$ and a 
linear dispersion relation $\omega = v |\bk|$ for $\bk \to 0$.
Within the local elasticity model (without long-range interactions) the linear dispersion 
relation can only emerge when an external source, such as an in-plane stress, breaks the 
rotational symmetry explicitly.
In the harmonic approximation, this follows from the fact that in a 
rotationally-invariant, unstressed membrane, a Lagrangian term proportional to 
$(\pa_{\alpha}\bh)^{2}/2$ cannot appear individually, but only together with in-plane 
terms in an overall coupling $\sigma_{A} U_{\alpha \alpha}$ to the strain tensor 
$U_{\alpha \alpha} = (\pa_{\alpha}\br \cdot \pa_{\alpha}\br -2)/2$.
The contribution $\sigma_{A} U_{\alpha \alpha}$ represents a coupling to the change of 
the total area, rather than the in-plane area~\cite{ramirez_prb_2017}, and thus it is 
allowed without breaking the symmetry.
However, $U_{\alpha \alpha}$ contains a term linear in $\pa_{\alpha}u_{\alpha}$ and, 
thus, shifts the equilibrium configuration at which the energy must be expanded.
After expansion at the true energy minimum, the sound velocity term must 
disappear.
Indeed, a term linear in $\sigma_{A} U_{\alpha \alpha}$ can always be fully removed from 
the action by a change of variables $\br \to \zeta \br$~\cite{aronovitz_jpf_1989}, which 
is allowed for a membrane with free boundaries.
After $\zeta$ is chosen in such way that terms linear in $\pa_{\alpha}u_{\alpha}$ 
disappear, the entire operator $U_{\alpha \alpha}$ drops from the action, showing that 
the inverse Green function behaves as $G^{(0)-1}_{ij}(0, \bk) \propto k^{4}$.
Beyond the harmonic approximation, the perturbative corrections can be assembled in the 
effective potential $\Gamma$, generating functional of one-particle irreducible 
correlation functions, which, by Ward identities, has the same symmetry of the 
action~\cite{zinn-justin_qft}.
Because phonon excitations are gapless, the interacting Green function $G^{-1}_{ij}(0, 
\bk)$ can vanish slower than $k^{4}$ as a result of singular diagrammatic contributions 
which generate a non-analytic dependence on $\bk$.
However, the rotational symmetry forbids terms regular in $\bk$ and proportional to 
$k^{2}$.
For example, the first-order perturbative correction for classical membranes in two 
dimensions~\cite{nelson_jpf_1987, katsnelson_graphene} consists in a diagram suppressed by 
an overall factor $k^{4}$ but multiplied by a singular term $1/k^{2}$ arising from the 
loop integration.
Although formally the contribution vanishes as $k^{2}$, its origin is different from a 
regular contribution directly proportional to $k^{2}$.
Furthermore, the singular first-order term proportional to $k^{2}$ is, in fact, the first 
contribution to an infrared-divergent series which requires a resummation for example by 
the self-consistent screening approximation~\cite{gazit_pre_2009, le-doussal_aop_2018}, or 
the $\varepsilon$-expansion~\cite{aronovitz_prl_1988, mauri_npb_2020, metayer_pre_2022}.
After resummation, the interacting correlation function for classical membranes can be 
shown to behave as $k^{4 - \eta_{*}}$ where $\eta_{*}$ is an universal 
exponent~\cite{aronovitz_prl_1988, gazit_pre_2009, le-doussal_aop_2018}.
The value of $\eta_{*}$ has been computed by several complementary 
techniques~\cite{aronovitz_prl_1988, kownacki_pre_2009, gazit_pre_2009, coquand_pre_2016, 
le-doussal_aop_2018, saykin_aop_2020, mauri_npb_2020, coquand_pre_2020, metayer_pre_2022, 
pikelner_arxiv_2021} (see also references in~\cite{mauri_npb_2020}) and, despite some 
scatter between different methods, is usually found to be approximately $\eta_{*} \simeq 
0.8$.
As a result, the full interacting Green function $G^{-1}$ vanishes faster than $k^{2}$.

The same conclusions hold in presence of the linearized rotational 
invariance~\eqref{rotation}, which forces the action to depend on $(\pa_{\alpha}\bh)^{2}$ 
only via the linearized strain tensor $u_{\alpha \beta}$.
(See Refs.~\cite{aronovitz_prl_1988, guitter_jpf_1989, guitter_jpf_1990} for a discussion 
of Ward identities).
Terms linear in $u_{\alpha \alpha}$ can be fully removed by a change of variables 
$u_{\alpha} \to u_{\alpha} + \varepsilon x_{\alpha}$, a shift which is automatically 
performed when integrating over zero modes in the fixed-stress ensemble (see 
Sec.~\ref{sec:integration-in-plane}).

The prediction of a self-energy correction $\Sigma(\bk) \approx k^{2} \Lambda^{2}$, 
derived in Ref.~\cite{amorim_prb_2014}, resulted from a theory in which the in-plane 
kinetic energy was kept but the strain tensor was approximated.
This approximation breaks explicitly both the full and the linearized rotational 
symmetry, leading to a result inconsistent with the Ward identities.

\subsection{Average $\langle \hat{\bh}(\bk) \hat{\bh}(-\bk)\rangle$ at $T = 0$}

The quantum-mechanical average $\langle \hat{\bh}(\bk) \hat{\bh}(-\bk)\rangle$ is given by 
the integral $\int {\rm d}\omega/(2\pi) G(\omega, \bk)$ over all frequencies.
The scaling relations~\eqref{RG-GCI} and the one-loop approximation imply that $\langle 
\hat{\bh}(\bk) \hat{\bh}(-\bk)\rangle \approx 1/(z^{1/2}(k)k^{2})$.

\subsection{Anomalous Hooke's law at $T = 0$}

The average strain of the membrane $\langle \pa_{\alpha} u_{\beta} \rangle = v 
\delta_{\alpha \beta}$ can be computed from the thermodynamic relation $v = 
-\frac{1}{2}(\pa \Gr/\pa \sigma)$.
Thus, it is the sum $v = \sigma/(2B) + \Delta v$ of the Hookean term and the 
fluctuation part $\Delta v = - \frac{1}{2}(\pa \Delta \Gr/\pa \sigma) = - \langle 
(\pa_{\alpha}\bh)^{2} \rangle/4$.
The RG equation~\eqref{RG-Gibbs-T} for the Gibbs free energy implies that, at zero 
temperature
\begin{equation}\label{RG-Delta_v}
\begin{split}
\bigg[& \Lambda \frac{\pa}{\pa \Lambda} + \beta(Y)\frac{\pa}{\pa Y} + 2 \eta \sigma 
\frac{\pa}{\pa \sigma} + \eta \bigg] \Delta v \\
& = -\frac{1}{2} (\bar{b}_{1}(Y) \Lambda^{2} + \bar{b}_{2}(Y) \sigma)~.
\end{split}
\end{equation}
The inhomogeneous coefficients $\bar{b}_{1}(Y)$ and $\bar{b}_{2}(Y)$ are non-zero 
already in the non-interacting model, because the Gibbs free energy at $T = 0$ 
(equivalent to the zero-point ground state energy) of free flexural phonons is
\begin{equation}\label{Gibbs-divergences-0}
\begin{split}
\Delta \Gr_{0} & = \frac{d_{c}}{2} \int \frac{{\rm d}^{2}k}{(2\pi)^{2}} \sqrt{
k^{4} + \sigma k^{2}} \\
& = C_{0} + \frac{d_{c}}{16 \pi} \sigma \Lambda^{2} - \frac{d_{c}}{32 \pi} \sigma^{2} 
\ln \Lambda +\text{finite}~,
\end{split}
\end{equation}
and already contains divergences for $\Lambda \to \infty$.
Matching Eq.~\eqref{Gibbs-divergences-0} with the RG equation~\eqref{RG-Gibbs-0}, we 
deduce $\bar{b}_{1}(Y) = d_{c}/(8\pi) + {\rm O}(Y)$ and $\bar{b}_{2}(Y) = - 
d_{c}/(16 \pi) + {\rm O}(Y)$.
Taking into account that the expansion of $\beta$ and $\eta$ start, respectively, 
at orders $Y^{2}$ and at order $Y$, it can be checked that the general solution of 
the RG equation~\eqref{RG-Delta_v} order by order in $Y$ has the general structure
\begin{equation}\label{v-logarithms}
v = v_{0} + \frac{\sigma}{2B} + \sum_{k=0}^{\infty} \sum_{\ell=0}^{k+1} a_{k\ell} 
\sigma Y^{k} \lt(\ln(\Lambda^{2}/\sigma)\rt)^{\ell}~.
\end{equation}
The first term, $v_{0}$, represents the average strain at zero imposed stress, and the 
second two terms describe the response to external tension.

By solving the RG equation~\eqref{RG-Delta_v} in the leading-logarithm 
approximation~\cite{zinn-justin_qft} (keeping only the most singular terms, with $\ell = 
k+1$), we find
\begin{equation}\label{anomalous-Hooke-T=0}
\begin{split}
v - v_{0} & = \frac{\sigma}{2B} + \frac{4 \sigma}{3 Y} 
\big[(z(\sigma^{\frac{1}{2}}))^{\frac{d_{c}}{4}} - 1\big]~,
\end{split}
\end{equation}
where $z(\sigma^{1/2})$ is the running amplitude defined in 
Eq.~\eqref{1L-renormalizations}, evaluated at scale $\Lambda' = \sigma^{1/2}$.

Eq.~\eqref{anomalous-Hooke-T=0} is consistent, at the leading logarithm level, with 
results obtained by other methods in Ref.~\cite{guitter_jpf_1990, burmistrov_prb_2016}, 
and shows that, even at $T = 0$, the stress-strain relation is anomalous.
Due to quantum fluctuations of flexural degrees of freedom, the macroscopic bulk modulus 
$B_{\rm eff} = \frac{1}{2} \pa \sigma/\pa v$ is not a constant, but a slowly-varying 
function of the applied tension:
\begin{equation}\label{B-eff}
\frac{1}{B_{\rm eff}(\sigma)} \approx \frac{1}{B} + \frac{8}{3 Y} \bigg[\bigg(1 + 
\frac{3(d_{c}+6) Y}{256 \pi} \ln \frac{\Lambda^{2}}{\sigma}\bigg)^{\frac{d_{c}}{d_{c} 
+ 6}}-1\bigg]~.
\end{equation}
(In Eq.~\eqref{B-eff} we neglected a contribution from the derivative of 
$z(\sigma^{1/2})$, which does not contribute to the leading-logarithmic singularities).
In the limit of zero tension $\sigma \to 0$, the bulk modulus vanishes as $B_{\rm 
eff}(\sigma) \approx [\ln (\Lambda^{2}/\sigma)]^{-d_{c}/(d_{c}+6)}$.

The physical origin of this singularity is the same which gives rise to the anomalous 
Hooke response in classical thermally-fluctuating membranes~\cite{guitter_jpf_1989, 
gornyi_2dmater_2017, nicholl_prl_2017, burmistrov_aop_2018}: for small $\sigma$ the 
dominant effect of the applied tension is not a stretching of the interatomic distance, 
but rather a "flattening" of the distribution of out-of-plane flexural fluctuations.
The singularity, in particular, is much weaker than the power-law-divergent anomalous 
Hooke's law characteristic of classical thermal fluctuations, derived in 
Refs.~\cite{guitter_jpf_1989, gornyi_2dmater_2017, burmistrov_aop_2018}.

Similarly to the momentum-dependence of $\kappar(k)$ and $\Yr(k)$, the stress-strain 
relation does not exhibit any characteristic crossover, reflecting the simple form of the 
RG beta function $\beta \propto Y^{2}$.
The strain response is thus given by a linear Hooke law corrected by logarithmic factors, 
for all values of the tension.
Eqs.~\eqref{anomalous-Hooke-T=0} and~\eqref{B-eff} break down, however, when $\sigma$ 
reaches the cutoff scale $\sigma \approx \Lambda^{2}$. In conventional units, 
this corresponds to a value of the tension $\sigmar = \kappar \Lambda^{2}$.
The contribution of quantized out-of-plane fluctuations dominates over the regular 
Hookean 
response only at exponentially suppressed values of the tension $\sigma \ll \sigma_{1} 
\approx \Lambda^{2} {\rm e}^{-2/g_{0}}$, which for parameters characteristic of graphene 
and two-dimensional materials (see Sec.~\ref{sec:graphene}) corresponds to an 
unphysically 
small stress.
However, the logarithmic corrections induced by out-of-plane motion are nonzero also at 
much larger values of the stress, and do not present any qualitative change of behavior 
near $\sigma = \sigma_{1}$.
A crossover is expected, instead, for membranes of finite size, since the linear 
dimension of the system then provides an independent scale.
The infinite-size predictions can be assumed to remain valid for $\sigma \gg l^{-2}$, 
where $l$ is the characteristic linear size.

\subsection{Consequences of renormalizability on low-temperature thermodynamics}

Differentiating Eq.~\eqref{RG-Gibbs-T} with respect to the temperature $T$, annihilates 
the inhomogeneous terms $b_{0} + b_{1} \sigma + b_{2} \sigma^{2}/2$, which are 
temperature-independent.
As a result we find a homogeneous renormalization group equation for the  entropy per 
unit area $S = - \pa \Gr/\pa T|_{\sigma}$:
\begin{equation}
\bigg[\Lambda\frac{\pa}{\pa \Lambda} + \beta(Y) \frac{\pa}{\pa Y} + 2 \eta 
\sigma \frac{\pa}{\pa \sigma} + \eta T\frac{\pa}{\pa T}\bigg] S = 0~,
\end{equation}
valid in the limit of small tension $\sigma \ll \Lambda^{2}$ and small temperature $T 
\ll \Lambda^{2}$.
By further differentiation with respect to $T$ and to $\sigma$ we find RG equations 
for the specific heat at constant tension $C = T \pa S/\pa T|_{\sigma}$ and for the 
thermal expansion coefficient $\alpha = 2\pa v/\pa T|_{\sigma} =\pa S/\pa \sigma|_{T}$:
\begin{equation}
\bigg[\Lambda\frac{\pa}{\pa \Lambda} + \beta(Y) \frac{\pa}{\pa Y} + 2 \eta 
\sigma \frac{\pa}{\pa \sigma} + \eta T \frac{\pa}{\pa T}\bigg]C = 0~,
\end{equation}
\begin{equation}
\bigg[\Lambda\frac{\pa}{\pa \Lambda} + \beta(Y) \frac{\pa}{\pa Y} + 2 \eta 
\sigma \frac{\pa}{\pa \sigma} + \eta T \frac{\pa}{\pa T} + 2 \eta\bigg]\alpha = 0 
~.
\end{equation}
By using the standard method of characteristics~\cite{zinn-justin_qft}, the solutions can 
be written as
\begin{equation}\label{solution-characteristics}
\begin{split}
S(Y, \sigma, T, \Lambda) & = S(y(\Lambda'), z^{-1}\sigma, z^{-1/2}T, 
\Lambda')\\
C(Y, \sigma, T, \Lambda) & = C(y(\Lambda'), z^{-1}\sigma, z^{-1/2}T, 
\Lambda')\\
\alpha(Y, \sigma, T, \Lambda) & = z^{-1}\alpha(y(\Lambda'), z^{-1}\sigma, 
z^{-1/2}T , \Lambda')~,
\end{split}
\end{equation}
where $y(\Lambda')$ and $z = z(\Lambda')$ are the zero-temperature running couplings 
introduced in Sec.~\ref{sec:two-point-correlations}.

Some general consequences of the RG equations, however, become more manifest if the 
solutions are expressed in another well-known form~\cite{gell-mann_pr_1954}.
By rewriting the definitions~\eqref{running-couplings} of the flow of running coupligs in 
the integral form
\begin{equation}\label{varphi&f}
\ln \frac{\Lambda'}{\Lambda} = \varphi(y(\Lambda')) - \varphi(Y)~, \qquad z(\Lambda') = 
\frac{{\rm e}^{f(y(\Lambda'))}}{{\rm e}^{f(Y)}}~, 
\end{equation}
\begin{equation}
\varphi(x) = \int^{x}\frac{{\rm d}u}{\beta(u)}~, \qquad f(x) = -2 \int^{x} {\rm d}u  
\frac{\eta(u)}{\beta(u)}~.
\end{equation}
it can be checked that the dimensionless quantities 
\begin{equation}\label{RG-invariants}
\begin{split}
x_{1} & = \ln \frac{\Lambda^{2}}{T} -\frac{1}{2} f(Y) - 2\varphi(Y)~,\\
x_{2} & =  \frac{\sigma}{T} {\rm e}^{\frac{1}{2} f(Y)}~,
\end{split}
\end{equation}
are RG-invariant (they do not change under the replacements $\Lambda \to \Lambda'$, $Y 
\to y(\Lambda')$, $T \to z^{-1/2}(\Lambda') T$, $\sigma \to z^{-1}(\Lambda')\sigma$).

Taking into account that $\alpha$ is dimensionless, while $S$ and $C$ have the dimension 
of an inverse area, the scaling relations can then be written in the form, equivalent to 
Eq.~\eqref{solution-characteristics},
\begin{equation}\label{solution-universality}
\begin{split}
S(Y, \sigma, T, \Lambda) & = T {\rm e}^{\frac{1}{2}f(Y)} L(x_{1}, x_{2})~,\\
C(Y, \sigma, T, \Lambda) & = T {\rm e}^{\frac{1}{2}f(Y)} M(x_{1}, x_{2})~,\\
\alpha(Y, \sigma, T, \Lambda) & = {\rm e}^{f(Y)} N(x_{1}, x_{2})~,\\ 
\end{split}
\end{equation}
where $L$, $M$, and $N$ are fixed functions of two parameters.
The thermodynamic relations between $S$, $C$, and $\alpha$ imply
\begin{equation}
\begin{split}
M(x_{1}, x_{2}) & = \bigg[1 - \frac{\pa}{\pa x_{1}} - x_{2} \frac{\pa}{\pa 
x_{2}}\bigg]L(x_{1}, x_{2})~,\\
N(x_{1}, x_{2}) & = \frac{\pa L(x_{1}, x_{2})}{\pa x_{2}}~.
\end{split}
\end{equation}
The detailed form of the functions $L$, $M$, and $N$ is not fixed by the scaling 
relations, but requires a full solution of the problem, including an analysis of the 
long-wavelength degrees of freedom dominated by classical thermal fluctuations.
However Eqs.~\eqref{solution-universality}, which are general consequences of the 
renormalizability of the zero-temperature theory, already have a predictive content, even 
without a full solution of the problem.
They imply that, in the region $\sigma  \ll \Lambda^{2}$, $T\ll \Lambda^{2}$ thermodynamic 
quantities depend on the microscopic material parameters $\Lambda$ and $Y$ only via 
overall scale factors independent of $\sigma$ and $T$.

For example, the form of the thermal expansion coefficient at zero tension
\begin{equation}\label{alpha-renormalizability}
\alpha = {\rm e}^{f(Y)} N(\ln (\Lambda^{2}/T) - f(Y)/2 - 2\varphi(Y), 0)
\end{equation}
implies that, in a logarithmic plot of $\ln \alpha$ vs. $\ln T$, curves corresponding to 
different materials must have the same shape and differ only by rigid shifts along 
horizontal and vertical Cartesian axes.

These universality properties express, in the thermodynamical behavior, a standard 
consequence of renormalizability~\cite{gell-mann_pr_1954}.

\subsection{Finite-temperature thermodynamics of an unstressed membrane: thermal 
expansion coefficient}

A more detailed prediction of the temperature dependence of thermodynamic quantities 
requires a complete theory of all degrees of freedom, from short-wavelength modes, frozen 
by quantization, to long-wavelength modes, controlled by thermal fluctuations.

To derive explicit expressions we use a combination of the scaling 
relations~\eqref{solution-characteristics} with approximations analogue to those 
described in Ref.~\cite{burmistrov_prb_2016}.

The theories of Refs.~\cite{coquand_pre_2016, burmistrov_prb_2016} indicate that, for 
temperatures much smaller than the Debye temperature $\Tr_{\rm D}$, correlation functions 
exhibit a double crossover between different regimes.
For large momenta $|\bk| \gg q_{T}$ flexural modes have fluctuations of zero-point 
character.
In an intermediate window of length scales $q_{\rm G} \ll |\bk| \ll q_{T}$ the system is 
expected to exhibit weakly-coupled harmonic fluctuations and a classical statistical 
distribution.
Finally in the long-wavelength region $|\bk| \ll q_{\rm G}$, fluctuations become strongly 
anharmonic and are controlled by the interacting fixed point characteristic of classical 
membranes~\cite{nelson_statistical, aronovitz_prl_1988, aronovitz_jpf_1989, 
kownacki_pre_2009, le-doussal_aop_2018, shankar_pre_2021}.

The crossover scale $q_{T}$ separating zero-point from thermally activated regimes, can be 
estimated~\cite{burmistrov_prb_2016}, as the wavelength at which the zero-temperature 
inverse Green function $G^{-1}_{ij}(0, \bk) \approx z(k)k^{4}$ becomes of the order of 
$T^{2}$:
\begin{equation}\label{qT}
z(q_{T})q_{T}^{4} \simeq T^{2}~.
\end{equation}
For a fully classical membrane with bending rigidity $\kappar_{1}$, Young modulus 
$\Yr_{1}$, and temperature $\Tr_{1}$, the Ginzburg momentum $q_{\rm G}$ at which harmonic 
fluctuations cross over to strongly-coupled nonlinear fluctuations 
is~\cite{nelson_statistical, aronovitz_jpf_1989, le-doussal_aop_2018, 
burmistrov_aop_2018, shankar_pre_2021} $q_{\rm G} \simeq (3 k_{\rm B}\Tr_{1} \Yr_{1}/(16 
\pi \kappar_{1}^{2}))^{1/2}$.
In the quantum problem it can be assumed that the same crossover criterion remains valid, 
with $\Tr_{1} = \Tr$ and that $\kappar_{1} = \kappar_{\rm r}(q_{T})$, $\Yr_{1} = \Yr_{\rm 
r}(q_{T})$ are the renormalized parameters~\eqref{kappaR},~\eqref{YR}, corrected by 
zero-point anharmonic effects, evaluated at the renormalization scale 
$q_{T}$~\cite{burmistrov_prb_2016}.
In rescaled units, the corresponding crossover scale is
\begin{equation}\label{qG}
q_{G}^{2} \simeq \frac{3 T y(q_{T})}{16 \pi(z(q_{T}))^{1/2}}~.
\end{equation}
With characteristic parameters of graphene (see Sec.~\ref{sec:graphene}), it can be 
verified that $(q_{G}/q_{T})^{2}  = 3 y(q_{T})/16 \pi$ is small, confirming the 
consistency of a region $q_{\rm G} \ll |\bk| \ll q_{T}$.

By using Eq.~\eqref{solution-characteristics}, we can estimate the thermal expansion of 
the quantum membrane as
\begin{equation}
\alpha(Y, T, \Lambda) = (z(q_{T}))^{-1} \alpha(y(q_{T}), z(q_{T})^{-1/2}T, q_{T})~.
\end{equation}
In principle, the zero-temperature RG flow remains valid only as far as $\Lambda' \gg 
z^{-1/2}(\Lambda') T$, but, in a first approximation, it is justified to set directly 
$\Lambda' = q_{T} = z^{-1/2}(q_{T}) T$.

After the cutoff has been reduced from the microscopic scale to the thermal scale $q_{T}$, 
we can estimate $\alpha$ by neglecting quantum thermal effects and by identifying 
$\alpha(y(q_{T}), z^{-1/2}(q_{T})T, q_{T})$ with the thermal expansion coefficient of a 
classical membrane with the standard Hamiltonian~\cite{nelson_statistical, 
nelson_jpf_1987, aronovitz_jpf_1989, burmistrov_prb_2016, coquand_pre_2016}
\begin{equation}\label{H1-cl}
\begin{split}
{\cal H}_{\rm cl}[\bH_{1}(\bx)] & = \int {\rm d}^{2}x \bigg[\frac{\kappa_{\rm 
cl}}{2}(\pa^{2}\bH_{1})^{2} \\
& + \frac{Y_{\rm cl}}{8} (P^{T}_{\alpha \beta}(-\pa^{2})(\pa_{\alpha}\bH_{1}\cdot 
\pa_{\beta}\bH_{1}))^{2}\bigg]~.
\end{split}
\end{equation}
In terms of the discussion of Sec.~\ref{sec:RG-low-T}, this corresponds to approximating 
${\cal H}[\bH_{1}]$, the Hamiltonian for modes with zero Matsubara frequency, with 
Eq.~\eqref{H1-cl}, which is its tree-level approximation (without loop corrections).

In particular, we must consider a classical membrane with Young modulus 
$Y_{\rm cl} = y(q_{T})$, temperature $T_{\rm cl} = z^{-1/2}(q_{T})T$, bending rigidity 
$\kappa_{\rm cl} = 1$, and a large-momentum cutoff $\Lambda_{\rm cl} = q_{T}$.

Thermal fluctuations in classical statistical mechanics have been investigated 
extensively~\cite{nelson_statistical, nelson_jpf_1987, aronovitz_jpf_1989, 
burmistrov_aop_2018, le-doussal_aop_2018, mauri_npb_2020, saykin_aop_2020, 
coquand_pre_2020, shankar_pre_2021}.
The momentum-dependent correlation function $G_{ij}^{({\rm cl})}(\bk) =\langle h_{i}(\bk) 
h_{j}(-\bk)\rangle$ is predicted to behave as
\begin{equation}
G_{ij}^{({\rm cl})}(\bk) = \delta_{ij} \begin{cases}
                                        \frac{T_{\rm cl}}{\kappa_{\rm cl}k^{4}}~, & 
\text{for } q_{\rm G} \ll |\bk| \ll\Lambda_{\rm cl}\\
\frac{T_{\rm cl}}{\kappa_{\rm cl}k^{4 - \eta_{*}}q_{\rm G}^{\eta_{*}}} & \text{for }  
|\bk| \ll q_{\rm G}~,
                                       \end{cases}
\end{equation}
where $\eta_{*}$ is an universal exponent and $q_{G} = (3 T_{\rm cl} Y_{\rm cl}/(16 
\pi \kappa_{\rm cl}^{2}))^{1/2}$.

Calculating directly the extension factor via the relation
\begin{equation}
\begin{split}
\langle v_{\rm cl} \rangle & = -\frac{1}{4} \langle (\pa_{\alpha}\bh)^{2} \rangle = 
-\frac{1}{4} \int \frac{{\rm d}^{2}k}{(2\pi)^{2}} G_{ii}(\bk)\\
& \simeq -\frac{d_{c} T_{\rm cl}}{8 \pi \kappa_{\rm cl}} \bigg[\int_{0}^{q_{\rm G}} 
\frac{{\rm d}k}{q_{\rm G}^{\eta_{*}}k^{1-\eta_{*}}} + \int_{q_{\rm G}}^{\Lambda_{\rm 
cl}} \frac{{\rm d}k}{k}\bigg]\\
& = - \frac{d_{c}T_{\rm cl}}{8 \pi \kappa_{\rm cl}}\bigg[\frac{1}{\eta_{*}} + \ln 
\frac{\Lambda_{\rm cl}}{q_{\rm G}}\bigg]
\end{split}
\end{equation}
and differentiating with respect to $T_{\rm cl}$ at $\Lambda_{\rm cl}$, $Y_{\rm cl}$ and 
$\kappa_{\rm cl}$ fixed we find the expression for the thermal expansion 
coefficient~\cite{de-andres_prb_2012} 
\begin{equation}
\alpha_{\rm cl} = 2 \frac{\pa \langle v_{\rm cl}\rangle}{\pa T_{\rm cl}} = - 
\frac{d_{c}}{4 \pi \kappa_{\rm cl}} \lt[\frac{1}{\eta_{*}} - \frac{1}{2} + \ln 
\frac{\Lambda_{\rm cl}}{q_{\rm G}}\rt]~.
\end{equation}
Identifying, in rescaled units, $\alpha = (z(q_{T}))^{-1} \alpha_{\rm cl}$, and setting 
the effective classical parameters to the renormalized values we then find an expression 
for the thermal expansion coefficient of quantum membranes
\begin{equation}\label{alpha-solution}
\alpha = -\frac{d_{c}}{4 \pi z(q_{T})} \bigg[\frac{1}{\eta_{*}} - \frac{1}{2} + 
\frac{1}{2}\ln \bigg(\frac{16 \pi}{3 y(q_{T})}\bigg)\bigg]~.
\end{equation}
This expression, when $z(q_{T})$ and $y(q_{T})$ are replaced with the one-loop running 
couplings
\begin{equation}
y(q_{T}) = \frac{Y}{1 + \frac{3(d_{c}+6)Y}{128 \pi} \ln \frac{\Lambda}{q_{T}}}~, \quad 
z(q_{T}) = \lt(\frac{Y}{y(q_{T})}\rt)^{\theta}~,
\end{equation}
coincides with the result derived in Ref.~\cite{burmistrov_prb_2016}, up to a numerical 
factor.

The temperature dependence of $\alpha$ is entirely driven by the renormalization factors 
$y(q_{T})$ and $z(q_{T})$.
As a result, the thermal expansion coefficient is a slow, logarithmic function of $T$.
In the limit $T \to 0$, $\alpha$ tends to zero as
\begin{equation}\label{alpha-T}
\alpha_{T} \approx - \frac{d_{c}}{8\pi} \frac{\ln(\ln(\Lambda^{2}/T))}{[(g_{0}/2) \ln 
(\Lambda^{2}/T)]^{\theta}}~.
\end{equation}
As a remark, we note that, despite being approximate, the solution~\eqref{alpha-solution} 
is automatically consistent with the general form~\eqref{alpha-renormalizability}, as it 
is true for any expression of the type $\alpha = (z(q_{T}))^{-1} F(y(q_{T}))$, constructed 
via running couplings.

The approximations which lead to Eq.~\eqref{alpha-solution} are very natural.
We expect that the prediction of a nearly constant $\alpha$ at low temperature is exact 
and that Eq.~\eqref{alpha-T} captures, up to a numerical factor, the correct behavior in 
the limit $T \to 0$.

\subsection{Renormalization and third law of thermodynamics}

The fact that $\alpha$ vanishes in the zero-temperature limit is formally consistent with 
the requirement $\lim_{T \to 0} \alpha(T) = 0$, which is expected in view of the Maxwell 
relation
\begin{equation}
\alpha = \lt(\frac{\pa V}{\pa T}\rt)_{p} = -\lt(\frac{\pa S}{\pa p}\rt)_{T}~,
\end{equation}
and the third law of thermodynamics $\lim_{T\to 0} S(T, p) = 
0$~\cite{amorim_prb_2014, herrero_jcp_2016, landau_stat-phys-1}.
The logarithmic way in which the low temperature limit is realized at zero tension is, 
however, very unconventional~\cite{burmistrov_prb_2016}.
In fact, the existence of an anomalous behavior can be already anticipated by dimensional 
analysis.
The rescalings described in Sec.~\ref{sec:rescaled-units} show that, for $\sigma = 0$, 
the only dimensionful parameters in the theory~\eqref{action0-rescaled} are the 
temperature $T$ and the UV cutoff $\Lambda$.
If there were no ultraviolet divergences, the fact that $\alpha$ is dimensionless would 
have implied that $\alpha = \phi(Y)$, a temperature-independent result which is 
manifestly inconsistent with the limit $\lim_{T \to 0}\alpha(T) = 0$.
It is only the logarithmic correction due to UV divergences which introduces an explicit 
dependence on the UV cutoff scale $\Lambda$ and allows for a variation of $\alpha$ at low 
temperatures.
In presence of a nonzero tension $\sigma$, Ref.~\cite{burmistrov_prb_2016} predicted 
that the thermal expansion coefficient  vanishes in a faster way for $T \to 0$.

\section{Application to graphene}\label{sec:graphene}

To illustrate results, we consider the case of a monolayer graphene, using parameters 
$\rhor \simeq 7.6$ kg m$^{-2}$, $\lambdar \simeq 3.4$ eV \AA$^{-2}$, $\mur \simeq 9.3$ eV 
\AA$^{-2}$, $\Br \simeq 12.7$ eV\AA$^{-2}$, $\Yr \simeq $ 21.4 eV 
\AA$^{-2}$~\cite{los_prl_2016}, $\kappa \simeq 1.4$ eV~\cite{tisi_master-thesis}.
Setting the codimension $d_{c}$ to the physical value $d_{c} = 1$, we find that the bare 
value of the quantum coupling constant is small: $g_{0} \simeq 0.02$.
As a consequence, the one-loop approximations to the RG functions $\beta$ and $\eta$ are 
justified at all length scales of physical interest.
The smallness of $g_{0}$ is related physically to the fact that the mass of nuclei is 
much larger than the mass of electrons~\cite{kats_prb_2014, kats_prb_2014e}, and, thus, 
we expect it to be a general feature of most two-dimensional materials.

The ultraviolet cutoff $\Lambda$ is of the order of the inverse interatomic distance $a 
\simeq 1.42$ \AA.
We choose to identify $\Lambda$ with the "Debye radius" $\Lambda = (4 
\pi^{2}/3)^{1/4}a^{-1}$, defined by the condition that the phase space area $\pi 
\Lambda^{2}$ contains the same number of degrees of freedom of the hexagonal Brillouin 
zone of graphene.
With this estimate, the Debye temperature is approximately $\Tr_{\rm D} = \hbar 
(\kappar/\rhor)^{1/2}\Lambda^{2}/k_{\rm B} \simeq 750$ K. 

\begin{figure}
\centering 
\includegraphics[scale=1]{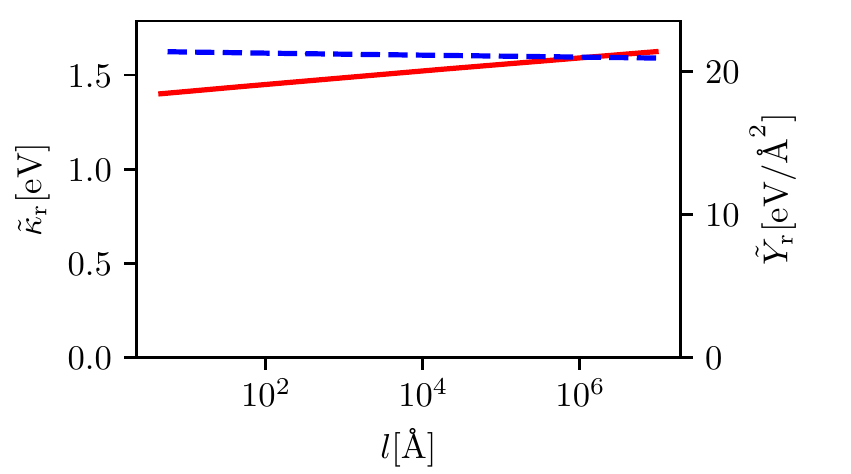}
\caption{\label{fig:kappar-Yr} Effective wavevector-dependent bending rigidity (red solid 
line) and Young modulus (blue dashed line) as a function of wavelength $l = 2\pi/k$ for a 
graphene membrane at $T = 0$, as described by Eqs.~\eqref{kappaR} and~\eqref{YR}.
In the infinite-wavelength limit $\kappar_{\rm r}(l)$ slowly diverges as $(\ln l)^{4/7}$ 
and $\Yr_{\rm r}(l)$ slowly vanishes as $(\ln l)^{-1/7}$.
}
\end{figure}

\begin{figure}
\centering 
\includegraphics[scale=1]{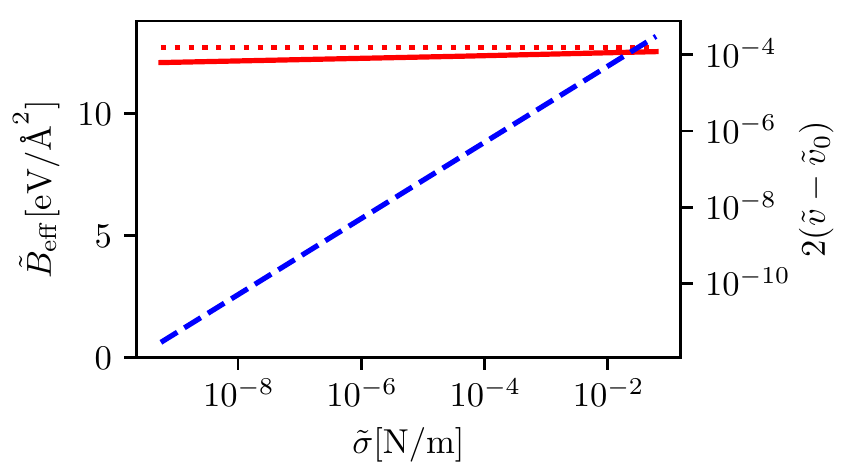}
\caption{\label{fig:B_sigma} Anomalous Hooke's law for a graphene membrane at $T = 0$.
The red solid line represents the macroscopic bulk modulus $\Br_{\rm eff} = \frac{1}{2} 
\pa \sigmar/\pa \tilde{v}$ as a function of the applied tension $\sigmar$, as described by 
Eq.~\eqref{B-eff}.
The red dotted line is constant as a function of the applied stress and identifies the 
microscopic bulk modulus $\Br \simeq 12.7$ eV \AA$^{-2}$ controlling the normal Hooke's 
law for a membrane constrained in two dimension (without quantum-mechanical out-of-plane 
fluctuations).
The strain induced by tension is represented by blue dashed lines.
The effective bulk modulus vanishes in the limit $\sigma \to 0$ as $\tilde{B}_{\rm 
eff}(\sigma) \approx (\ln (1/\sigma))^{-1/7}$.
The singularity, however, is very slow.
The prediction for the stress-strain relation breaks down when the tension is so large 
that the stress dominates over bending rigidity at all momentum scales up to the cutoff 
$\Lambda$.
Eq.~\eqref{B-eff} is thus valid for $\sigmar \ll \kappar \Lambda^{2} \simeq 40$ N/m.
}
\end{figure}

\begin{figure}
\centering
\includegraphics[scale=1]{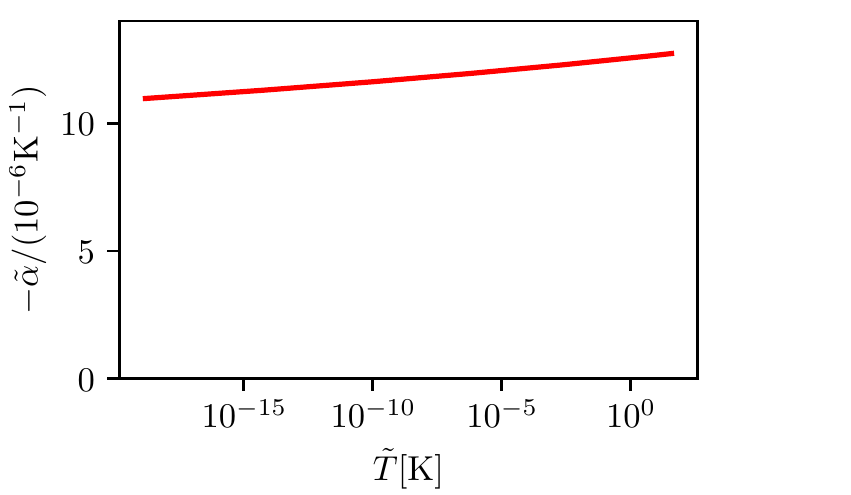}
\caption{\label{fig:alpha} Negative thermal expansion coefficient for an unstressed 
graphene membrane as a function of temperature (red solid line).
In the limit $T \to 0$, $\alphar \to 0$ as expected from the third law of thermodynamics, 
but the approach to zero is only logarithmic with $\Tr$.}
\end{figure}

The predictions discussed in Sec.~\ref{sec:results} are illustrated in 
figures~\ref{fig:kappar-Yr},~\ref{fig:B_sigma}, and~\ref{fig:alpha}.
In all cases, the renormalizations induced by quantum-mechanical fluctuations induce a 
slow, logarithmic behavior of statistical and thermodynamic quantities.

In particular, we find that the thermal expansion coefficient $\alphar$ at low 
temperatures is nearly constant over broad ranges of low temperature, with an order of 
magnitude $\alphar \approx -k_{\rm B}/(4\pi \kappar) \approx -5 \times 10^{-6}$ K$^{-1}$ 
up to a numerical factor of order unity, in agreement with a 
simple classical estimate~\cite{de-andres_prb_2012}.
The limit $\lim_{T \to 0} \alpha = 0$ is only approached logarithmically.

Let us estimate the characteristic crossover scales for flexural fluctuations.
At $\Tr = 300$~K, the crossover momentum $q_{T}$ between zero-point and classical 
fluctuations is approximately $q_{T} \simeq 0.9$~\AA$^{-1}$, corresponding to a 
wavelength $l_{T} \simeq 7.4$~\AA$^{-1}$.
The Ginzburg length separating harmonic and anharmonic classical regimes is, for $T \simeq 
300$~K, approximately $l_{G} = 2\pi/q_{G} \simeq 50$~\AA.
At different temperatures, the characteristic lengths $l_{T}$ and $l_{G}$ are both 
proportional to $\Tr^{-1/2}$, up to logarithmic factors.
This is a manifestation of the scale-invariance of the action, which is only weakly 
broken by logarithmic renormalization effects.

The prediction of a nearly constant $\alphar$ depends essentially on the fact that 
flexural phonon modes fluctuate in absence of an imposed stress and without binding 
forces (see Ref.~\cite{burmistrov_prb_2016} for an analysis on the role of tension and 
Refs.~\cite{de-andres_prb_2012, feng_small_2021} for discussions on the effects of a 
supporting substrate).

To conclude, we note that a more complete understanding of the thermodynamics of graphene 
samples requires a further analysis of the coupling between membrane fluctuations and 
Dirac electrons, which have been proposed to be at the origin of mechanical instabilities 
such as a spontaneous rippling~\cite{gazit_pre_2009, guinea_prb_2014}.
The role of electron fluctuations, however, is suppressed in insulating 2D materials such 
as hexagonal boron nitride.

\section{Summary and conclusions}

To summarize, we have analyzed the theory of a fluctuating quantum mechanical membrane 
within the framework of perturbative renormalization group techniques.
At zero temperature, the perturbative RG provides a systematic derivation of logarithmic 
singularities analyzed in earlier investigations by momentum-shell and by nonperturbative 
renormalization group techniques.
In the limit of a weakly applied external tension $\sigmar$, we recover the result that 
the stress-strain relation at $T = 0$ is singular: for $\sigmar \to 0$, the strain 
behaves as $\sigmar [\ln(1/\sigmar)]^{-1/7}$.

In the case of a small, but nonzero temperature, techniques of finite-size quantum field 
theory provide general scaling relations for thermodynamic quantities such as the entropy 
$S$, the specific heat $C$, and the thermal expansion coefficient $\alpha$ at vanishing 
or small external tension.
By an approximate solution of the theory, we derive that the negative thermal expansion 
coefficient of an unstressed crystalline membrane vanishes for $T \to 0$ as a logarithmic 
function of the temperature.

\section*{Acknowledgments}

This work was supported by the Dutch Research Council (NWO) via the Spinoza Prize.

\footnotetext[1]{We assume that the general power-counting principles for the effects of 
irrelevant interactions~\cite{zinn-justin_qft}, which are usually derived in the framework 
of Euclidean-invariant theories, remain valid in the nonrelativistic model considered 
here.
Specifically, we assume that the only effects of neglected irrelevant interactions at 
leading order for long wavelengths and low frequency is a renormalization of the value of 
the parameters $\rhor$, $\kappar$, $\lambdar$, and $\mur$, which become phenomenological 
quantities.
}

\footnotetext[2]{We note for example that the one-loop in-plane phonon tadpole graph, 
neglected in the effective theory~\eqref{action0} was shown~\cite{burmistrov_prb_2016} to 
give a contribution to the thermal expansion coefficient suppressed at low 
temperatures with respect to the flexural contributions considered here.}

\footnotetext[3]{A power-counting analysis restricted to the frequency-only part of the 
integrals shows that the $\omega$ integrations are all superficially convergent.
An application of the Weinberg theorem ensures the finiteness of 
integrations~\cite{arav_jhep_2019}.}

%

\end{document}